\documentclass[10pt,journal,compsoc]{IEEEtran}
\usepackage{amsmath,amsfonts}
\usepackage{algorithmic}
\usepackage{algorithm}
\usepackage{array}
\usepackage{textcomp}
\usepackage{stfloats}
\usepackage{url}
\usepackage{verbatim}
\usepackage{graphicx}
\usepackage{cite}
\usepackage{inconsolata}

\usepackage{amsmath,amsfonts, amssymb}
\usepackage{multirow}
\usepackage[ruled,lined,linesnumbered,vlined,algo2e]{algorithm2e}
\usepackage{graphicx}
\usepackage{textcomp}
\usepackage{xcolor}
\usepackage{framed}
\usepackage{listings}
\usepackage{bbding} 
\usepackage{color}
\usepackage{enumitem}
\usepackage{threeparttable}
\usepackage{xspace}
\usepackage{indentfirst}
\usepackage{booktabs}
\usepackage{subfig}
\usepackage{balance}
\usepackage{colortbl}
\usepackage{listings}
\usepackage{hyperref}

\newcommand{\lstbg}[3][0pt]{{\fboxsep#1\colorbox{#2}{\strut #3}}}
\definecolor{codegreen}{rgb}{0,0.6,0}
\lstdefinelanguage{diff}{
	frame=single,
        language=Java,
	basicstyle=\ttfamily\scriptsize\bfseries,
	morecomment=[f][\color{red}]{---}, 
	morecomment=[f][\color{codegreen}]{+++},
	morecomment=[f][\lstbg{red!20}]{-},
	morecomment=[f][\lstbg{green!20}]{+},
	morecomment=[f][\color{blue}]{@@},
        emphstyle=\bfseries\color{Rhodamine},
        commentstyle=\itshape\color{black!50!white}
}

\usepackage{ulem}

\usepackage[ruled,lined,linesnumbered,vlined,algo2e]{algorithm2e}

\definecolor{codegray}{rgb}{0.5,0.5,0.5}

\SetCommentSty{mycommfont}

\newcommand{\tool}{\textsf{VAScanner}\xspace}
\newcommand{\vulroot}{vulnerable root methods\xspace}
\newcommand{\vulapi}{vulnerable APIs\xspace}

\newcommand{\revise}[1]{\textcolor{black}{#1}}

\usepackage{pifont}
\usepackage{cleveref}

\begin{document}

\title{Does the Vulnerability Threaten Our Projects? Automated Vulnerable API Detection for Third-Party Libraries}

\author{Fangyuan Zhang,
        Lingling Fan*,
        Sen Chen, 
        Miaoying Cai,
        Sihan Xu,
        and~Lida Zhao
        % and~Zheli Liu
\IEEEcompsocitemizethanks{
	\IEEEcompsocthanksitem
    Fangyuan Zhang and Miaoying Cai are with DISSec, NDST, College of Computer Science, Nankai University, China. Emails: \{fangyuanzhang, miaoyingcai\}@mail.nankai.edu.cn.
    Lingling Fan (Corresponding author) and Sihan Xu are with DISSec, NDST, College of Cyber Science, Nankai University, China. Emails: \{linglingfan, xusihan\}@nankai.edu.cn.
	Sen Chen is with the College of Intelligence and Computing, Tianjin University, China. Email: senchen@tju.edu.cn. 
	Lida Zhao is with School of Computer Science and Engineering, Nanyang Technological University. Email: LIDA001@e.ntu.edu.sg.
}}

% The paper headers
\markboth{Journal of \LaTeX\ Class Files,~Vol.~XX, No.~XX, XX~2024}%
{Zhang \MakeLowercase{\textit{et al.}}: Does the Vulnerability Threaten Our Projects? Automated Vulnerable API Detection for Third-Party Libraries}

\IEEEtitleabstractindextext{
\begin{abstract}
Developers usually use third-party libraries (TPLs) to facilitate the development of their projects to avoid reinventing the wheels, however, the vulnerable TPLs indeed cause severe security threats. The majority of existing research only considered whether projects used vulnerable TPLs but neglected whether the vulnerable code of the TPLs was indeed used by the projects, which inevitably results in false positives and further requires additional patching efforts and maintenance costs (e.g., dependency conflict issues after version upgrades).
 
To mitigate such a problem,
we propose \tool, which can effectively identify \vulroot causing vulnerabilities in TPLs and further identify all \vulapi of TPLs used by Java projects. Specifically, we first collect the initial patch methods from the patch commits and extract accurate patch methods by employing a patch-unrelated \revise{sifting} mechanism, then we further identify the \vulroot for each vulnerability by employing an augmentation mechanism.
Based on them, we leverage backward call graph analysis to identify all \vulapi for each vulnerable TPL version and construct a database consisting of \revise{90,749 (2,410,779 with library versions)} \vulapi \revise{with 1.45\% false positive proportion with a 95\% confidence interval (CI) of [1.31\%, 1.59\%]} from {362} TPLs with {14,775} versions. The database serves as a reference database to help developers detect \vulapi of TPLs used by projects. Our experiments show {\tool eliminates {5.78\%} false positives and {2.16\%} false negatives owing to the proposed \revise{sifting} and augmentation mechanisms.}
{Besides, it outperforms the state-of-the-art method-level vulnerability detection tool {in analyzing direct dependencies}, Eclipse Steady, achieving more effective detection {of vulnerable APIs}.}
Furthermore, to investigate the real impact of vulnerabilities on real open-source projects, we exploit \tool to conduct a large-scale analysis on {3,147} projects {that depend on vulnerable TPLs}, and find only \revise{21.51\% of projects (with 1.83\% false positive proportion and a 95\% CI of [0.71\%, 4.61\%])} were threatened through \vulapi, {demonstrating that \tool can potentially reduce false positives significantly}.
\end{abstract}

\begin{IEEEkeywords}
Vulnerability Detection, Software Composition Analysis, Static Analysis
\end{IEEEkeywords}
}

\maketitle

\IEEEdisplaynontitleabstractindextext
\IEEEpeerreviewmaketitle

\IEEEraisesectionheading{\section{Introduction}\label{sec:introduction}}
\IEEEPARstart{J}{ava} developers frequently incorporate third-party libraries (TPLs) to speed up software development.
However, the utilization of TPLs may introduce security threats~\cite{zhan2021atvhunter,zhan2021research}. According to an open-source security and risk analysis report released by Synopsys~\cite{oss_security_report}, 97\% of the 2,409 codebases contained open-source components, and 81\% of them contained at least one known vulnerability.
To mitigate such a severe problem, software composition analysis (SCA)~\cite{SCAtool, owasp, ponta2018beyond, dependabot, ossindex, snyk, blackduck,sourceclear, ws, zhao2023software} is typically used to identify vulnerable TPLs. A couple of SCA tools have been suggested including Eclipse Steady~\cite{steady-page}, Dependabot~\cite{dependabot}, OSSIndex~\cite{ossindex}, OWASP Dependency Check~\cite{owasp}, etc.

However, from the detection side, nearly all SCA tools can only determine whether vulnerable TPLs are depended on by projects, but cannot tell whether vulnerable {API}s are actually invoked, {resulting in false positives introduced by analysis at the library level.} From the patch side, vulnerabilities introduced by TPLs can have unpredictable effects on the developers' projects. Once the vulnerabilities are detected, updating to a new version is the most straightforward way.
However, it may cause dependency conflict issues~\cite{wang2018dependency,wang2020watchman,wang2019could,wang2021will,liu2022demystifying} and compatibility issues~\cite{zhang2023compatible,zhang2023has,zhang2023mitigating,yang2023compatibility}, 
which will require substantial maintenance costs. 
Consequently, it is imperative to precisely determine whether the project is threatened by known vulnerabilities. In other words, if the vulnerability has a real negative impact on the project in practice, developers can generate a patch immediately to avoid an exploit of the vulnerability. If the vulnerability has no effect on the project, the handling of vulnerable TPLs is not urgent and can be incorporated into the regular development cycle. Thus, the real impact analysis of vulnerable TPLs at the method level is urgently needed no matter from the perspective of detection or patching~\cite{pashchenko2018vulnerable}. 

As far as we know, Eclipse Steady~\cite{plate2015impact,ponta2018beyond,ponta2020detection} is the only {open-source} work that provides a forward reachability analysis at the fine-grained method level for users.
However, according to our analysis, we conclude the following deficiencies in Steady: 
\textit{(1) The inaccuracy of patch method extraction.} Steady considers the methods whose abstract syntax trees have been changed in patch commits as patch methods, however, patch-unrelated methods may exist in patch commits, leading to false positives.
\textit{(2) The incompleteness of vulnerable root method identification.} Steady obtains \vulroot directly from patch commits, however, some \vulroot may exist in the commits that are not recognized or marked as patch commits. 
The incomplete identification would cause false negatives of vulnerable paths.
\textit{(3) Low efficiency of vulnerable path analysis.} Steady conducts forward reachability analysis for each TPL with low efficiency due to complex dependency analysis.

Therefore, in this paper, we aim to address the aforementioned problems to evaluate the real impact of vulnerable TPLs on projects.
However, we are facing the following challenges:
\textit{(1)} How to extract accurate patch methods from patch commits?
As we all know, not all modified methods in a patch commit are patch methods. Therefore, we need to \revise{sift} patch-unrelated methods out on the patch commit, to extract precise patch methods.
\textit{(2)} How to obtain comprehensive and precise \vulroot from patch commits? Due to the incompleteness of patch commits provided~\cite{li2024patchfinder}, it is not comprehensive to only handle the patch commits.
\textit{(3)} How to accurately scan the vulnerable code of libraries in the projects with less resource overhead? To ensure fewer resources spent during scanning, we need a comprehensive set of detected \vulapi of known vulnerable TPLs.

To fill the gap, we propose \tool (\uline{\textbf{V}}ulnerable \uline{\textbf{A}}PI \uline{\textbf{Scanner}}), an effective vulnerable API detection approach, to assess the impact of OSS vulnerabilities in Java projects.
We first collect public patch commits based on the vulnerability knowledge database and map the changed source code files involved in patch commits with class files in TPLs.
{We collect diff methods from patch commits as initial patch methods and then \revise{sift} out patch-unrelated methods to extract accurate patch methods.
We propose an augmentation mechanism to identify \vulroot based on these patch methods.}
Then we perform backward call-graph analysis on \vulroot and construct a vulnerable API database mapping with the relation among the vulnerable library versions, CVEs, and \vulapi, \revise{which includes 90,749 unique \vulapi (2,410,779 with library versions)} from {362} TPLs with {14,775} vulnerable versions involving {502} CVEs.
Based on the results, developers can figure out whether vulnerable libraries need to be patched at this time and prioritize the patches, thereby reducing additional patching efforts and maintenance costs.

To demonstrate the effectiveness of \tool, we conducted comprehensive experiments. 
{We took an in-depth analysis of the patch-unrelated methods \revise{sifted} out by the patch-unrelated \revise{sifting} mechanism, \vulroot introduced by the augmentation mechanism, {and vulnerable APIs in the vulnerable API database}.}
Moreover, we summarized 5 patterns of added patch methods, to analyze the fixed intention of introducing them.
{Based on statistical results, we \revise{sifted} out {1,352} patch-unrelated methods {with 98.06\% precision} and augmented \revise{249} \vulroot which were absent in patch commits {with \revise{93.57}\% precision}. {And the vulnerable API database constructed by \tool contains a total of \revise{90,749 unique vulnerable APIs with a false positive proportion of 1.45\% and a 95\% CI of [1.31\%, 1.59\%].}}
Furthermore, to demonstrate the effectiveness of our novel mechanisms, we conducted an ablation study on \tool and \tool- {with different mechanisms, and the result shows \tool eliminates {5.78\%} false positives and {2.16\%} false negatives.}
{Subsequently, we compared \tool with the state-of-the-art tool, Eclipse Steady. The experimental results have shown that \tool outperforms Steady {in analyzing direct dependencies}, achieving more comprehensive method-level detection (\#Cases: {214 vs. 95}). Specifically, Steady (Avg time: 769s) exists 61.71\% false negatives, while \tool (Avg time: 353s) yielded 2.97\% false positives and 20.45\% false negatives.}
Besides, our large-scale analysis on {3,147} real-world projects shows that only \revise{21.51\% of projects (with 1.83\% false positive proportion and a 95\% CI of [0.71\%, 4.61\%])} were \revise{potentially} threatened by \vulapi of TPLs, indicating the effectiveness of \tool.

In summary, we make the following contributions:

\begin{itemize}
    \item We proposed \tool, an effective and efficient tool that can detect \vulapi from TPLs used by Java projects, reducing false positives of vulnerabilities.

    \item We proposed two mechanisms to achieve accurate and complete vulnerable API identification for vulnerable libraries, i.e., a \revise{sifting} mechanism to \revise{sift} out patch-unrelated methods and an augmentation mechanism to augment the \vulroot, which eliminates {5.78\%} false positives and {2.16\%} false negatives.
    
    \item We constructed a reusable database including \revise{90,749} \vulapi \revise{(2,410,779 with library versions) with 1.45\% false positive proportion with a 95\% CI of [1.31\%, 1.59\%]} based on the identification results of \tool, which assists in achieving more efficient vulnerability detection than forward reachability analysis.
    
    \item We compared \tool with the state-of-the-art tool, Eclipse Steady. The experimental result demonstrates that \tool achieves more effective method-level identification in {analyzing direct dependencies}.
\end{itemize}

\section{{Background \& Concepts}}
\subsection{{Background}}
\noindent \textbf{The Maven Ecosystem.}
{The Maven ecosystem~\cite{maven} plays a crucial role in the Java landscape. It contains nearly 2,000 repositories and over 37 million packages. Each maven package is distinctly identified by the combination of GroupId, ArtifactId, and Version (GAV). Maven provides a simple and consistent approach by utilizing the configuration file (pom.xml) to effectively manage project dependencies, streamline the build process, and facilitate release development. Furthermore, since a maven package can be utilized as a TPL by other projects, it can be considered a project as well as a Java TPL.}

\noindent \textbf{Vulnerable Libraries and the Associated Risks.}
{Vulnerable libraries are TPLs that contain vulnerabilities. Using vulnerable libraries introduces potential security risks to the projects. For instance, the Log4Shell vulnerability~\cite{log4shell} existed in Apache log4j, which is a widely used Java-based logging library, affecting numerous projects.}

\noindent \textbf{Software Composition Analysis.}
{Software Composition Analysis (SCA)~\cite{SCA_78} involves analyzing the libraries and identifying their vulnerabilities. Vulnerable library identification is a subset of SCA, which typically relies on hash comparisons or configuration files (e.g., pom.xml) to identify TPLs, and detect vulnerable libraries based on vulnerability databases (e.g., NVD~\cite{nvd}). Vulnerability reachability analysis focuses on determining whether there is a path from the software to the vulnerable code in TPLs. This analysis often uses forward call analysis to ascertain whether the software can access the vulnerable code within the libraries.}

\begin{figure}[t]
\centering
  \includegraphics[width=0.5\textwidth]{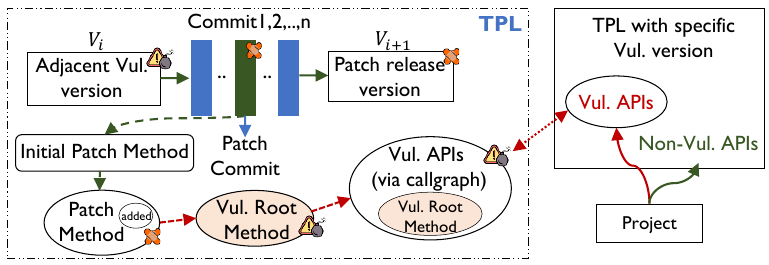}
  \caption{Illustration for terms used in the paper}
  \label{fig:concept}
\end{figure}

\subsection{Key Term Definition}
\label{sec:key-term}
{We introduce some key concepts or terms used in the paper to make it easy to understand, as illustrated in Figure~\ref{fig:concept}.}

\noindent \textbf{Adjacent Vul. version vs. Patch release version.} When an open-source TPL is affected by a vulnerability (also known as CVE), the vulnerability knowledge base usually gives the vulnerable version range of the TPL. 
``Patch release version'' means that it is the first release version to fix this vulnerability, i.e., $V_{i+1}$.
``Adjacent vulnerable version'' is the vulnerable version adjacent to the patch release version, i.e., $V_{i}$.
Patch commits used by developers to fix this vulnerability exist between these two versions.

\noindent \textbf{Initial Patch Method.} 
{Initial patch methods are the methods that have undergone code changes (i.e., added, deleted, or modified) in the patch commits.}

\noindent  \textbf{Patch Method.}
{Patch methods are methods that may be relevant to addressing vulnerabilities.}
Since not all initial patch methods play a role in patching, it is necessary to \revise{sift} out patch-unrelated methods (Section~\ref{sec:filter}) from the initial patch methods to generate precise patch methods. If a patch method is present only in the patch release version (i.e., $V_{i+1}$) and not in the adjacent vulnerable version (i.e., $V_{i}$), we consider it as an \textbf{added patch method}.

\noindent \textbf{Vul. Root Method.} 
\revise{Vulnerable root methods are those methods that are directly related to the vulnerability. Most of them are extracted from patch commits of vulnerabilities directly.}

\noindent \textbf{Vul. APIs.} 
{Vulnerable APIs are the methods that are directly or indirectly threatened by the vulnerability in the vulnerable TPL, including the \vulroot and the methods that directly/indirectly invoke \vulroot.}
For projects, APIs in TPLs are divided into 2 categories: \vulapi and non-\vulapi.

\subsection{Problem Definition}
As shown in Figure~\ref{fig:concept}, our goal is to identify all \vulapi for each vulnerable TPL version based on patch commits of CVEs and vulnerable root method identification and construct a database that maintains the mapping relation: \textit{vulnerable library versions} $\leftrightarrow$ \textit{CVEs} $\leftrightarrow $ \textit{\vulapi} (\textit{libV-CVE-Vul.API}), based on which we aim to detect whether the project invokes \vulapi of TPLs, to assess the real impact of OSS vulnerabilities on projects.

\begin{figure*}
\centering
  \includegraphics[width=1\textwidth]{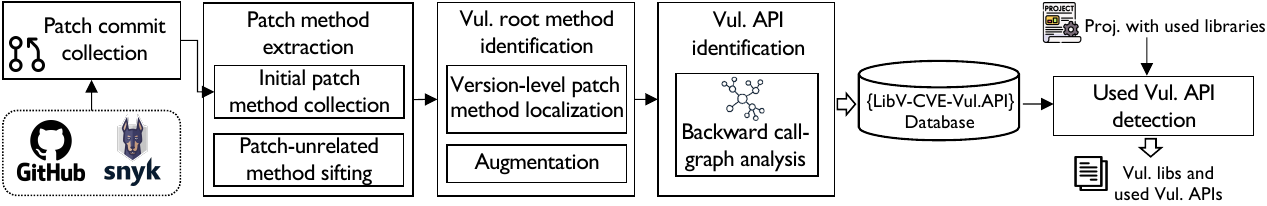}
  \caption{Overview of \tool}
  \label{fig:overview}
\end{figure*}

\section{Approach}
In this paper, we propose \tool to detect whether the projects are threatened by the \vulapi in TPLs. Figure~\ref{fig:overview} shows the overview of our approach, consisting of 4 components: (1) 
Patch method extraction, which collects initial patch methods from patch commits and \revise{sift}s out patch-unrelated methods to extract accurate patch methods.
(2) Vulnerable root method identification, which identifies \vulroot through {locating the patch methods at the version level and employing} an augmentation mechanism based on the extracted patch methods.
(3) Vulnerable API identification, which utilizes call-graph analysis to identify \vulapi for each library version, and constructs a database storing the mapping relations of vulnerable library versions (LibV), CVEs, and vulnerable {APIs} (Vul.API), presented as \textit{libV-CVE-Vul.API}. (4) Used vulnerable API detection, which detects the \vulapi in the libraries used by a given project.

\subsection{Patch Method Extraction}
This section describes the steps to extract the accurate \textit{patch methods}. Specifically, we first collect methods that have undergone code changes in the patch commits (i.e., \textit{initial patch methods}), and then \revise{sift} out patch-unrelated methods.

\subsubsection{Initial patch method collection}
To collect the methods related to patching vulnerabilities, we first need to obtain patch commits of each CVE.
Specifically, we collected vulnerabilities (identified by CVE ID) and their associated patch commits from Snyk Vulnerability DataBase~\cite{snyk_vulDB} and GitHub Advisory Database~\cite{gitadvisory}.
We chose them as the vulnerability data collection sources for two reasons:
(1) They maintain detailed information about CVEs and the corresponding patches, such as CVE ID, the vulnerable version ranges of TPLs, and patch-related links, which cover the CVE-related references provided by NVD.
Besides, for most fixed CVEs, the two databases provide patch commit references on GitHub~\cite{github}, which facilitates the collection and analysis of patch commits.
(2) They map CVEs to vulnerable libraries, allowing us to identify libraries with vulnerable versions based on CVE IDs.
Based on the two databases, we collected {2,640} CVEs and {1,551} affected {libraries} belonging to the Maven ecosystem.
We filtered out CVEs without patch commits in patch-related links and those where the affected libraries did not have patch release versions. Finally, we gathered 1,116 CVEs and 957 affected libraries to collect initial patch methods.

For each patch commit, we extracted code differences by using the abstract syntax tree (AST), as it can accurately identify real code changes and filter out irrelevant modifications like adding or deleting identical code, changing the position of methods, or adding blank lines. This approach is more effective and accurate than traditional code-based change extraction.
Specifically, we employed GumTree~\cite{falleri2014fine}, a tool for generating code differences in AST, to obtain valid changed methods in patch commits.
We first obtained the Java source code files before and after the commits based on the GitHub repository and used GumTree to generate the mappings between two ASTs.
{The identified code changes are divided into three types, i.e., insert, delete, and update. According to the tree structure representing methods in the AST, we got the signature of methods where different nodes were located. Finally, we obtained methods with valid code changes in the patch commits (i.e., \textit{initial patch methods}) with different change types (inserted, deleted, and modified).}
After filtering out CVEs whose patch commits involve languages other than Java (e.g., JavaScript), we consequently obtained the initial patch methods for {1,075} CVEs, {453} unique affected libraries and {1,350} patch commits.

\subsubsection{Patch-unrelated method \revise{sifting}}
\label{sec:filter}
Since not all initial patch methods are related to vulnerability fixing, we aim to \revise{sift} out patch-unrelated methods from initial patch methods, to obtain \textit{patch methods}.
{To achieve this, we initially extracted the changed (i.e., inserted, deleted, or updated) statements within each initial patch method. We then assessed whether these changed statements were unrelated to the patch. If all the changed statements within an initial patch method are patch-unrelated, the method will be \revise{sifted} out, otherwise, it is recognized as a patch method.}

{To achieve precise \revise{sifting} of patch-unrelated methods, we adopted a conservative strategy for identifying irrelevant statements. Specifically, we summarized three patterns of patch-unrelated statements:}
(1) \textbf{Debugging code statements}, such as \texttt{System.out.println(..)}, log-related function calls (e.g., \texttt{log.warn(..)}), and error handling statements which only changed the exception messages, i.e., \texttt{throws new xxException(..)};
{(2) \textbf{AST-equivalent statements after name normalization}.}
In detail, we initially collected the functions, class member variables, and formal parameters of functions that were solely renamed to generate a renaming set.
{We defined various renaming scenarios: when the function name changed but the function body remained unchanged, when a member variable merely altered its name but retained the same type and initialization, or when a formal parameter of a function only modified its name while maintaining the same type. In such instances, we categorized these functions, member variables, and formal parameters as being renamed.}
If a statement only includes modifications to the names of called functions, parameters of called functions, or the object of calling functions, we check whether the modified name exists in the renaming set. If it does, this statement is considered an AST-equivalent statement before and after the patch commit. Besides, if only the name of the assigned variable has been modified in an assignment statement (e.g., \texttt{A \uline{a} = foo()}), the statement will also be regarded as AST-equivalent;
(3) \textbf{Statements that solely compose the Getter/Setter functions},
such as \texttt{this.X = x}, \texttt{return X}, \texttt{return this} and \texttt{return this.X} (\texttt{X} is a class member variable). Note that we do not assert that the Getter/Setter functions are inherently patch-unrelated. Instead,
our goal is to identify and \revise{sift} out Getter/Setter functions that solely consist of those specific statements.

\smallskip
\subsection{Vulnerable Root Method Identification}
\label{sec:vulroot}
The patch methods are extracted based on patch commits, 
however, the patch release version or the adjacent vulnerable version of libraries (shown in \Cref{fig:concept}) may not contain the methods that were patched. Therefore, in this section, we aim to identify \textit{\textbf{\vulroot}} (denoted by $VulRoot$) by locating the patch methods at the version level instead of the commit level and augmenting them to obtain comprehensive \vulroot.

\subsubsection{Version-level patch method localization}
Since a commit only records a timestamped change to the current code in the repository, the changed methods in a single patch commit may not appear in the release versions of the {library}. For example, a library has several release versions {$V_1$, $V_2$, $V_3$, $V_4$..., $V_n$}, where $n$ is the number of versions, $V_2$ and $V_3$ are the vulnerable versions. 
There may be multiple commits between $V_3$ and $V_4$ aiming to patch the vulnerability in $V_3$, however, the changed methods in one commit might not be maintained in $V_4$ or exist in $V_3$, and should not be identified as a valid patch.
Therefore, we need to locate the patch methods at the version level to ensure they exist in the release versions. 

Specifically, We gathered all library versions from the Maven repository~\cite{maven} and extracted patch releases and adjacent vulnerable versions based on vulnerable version ranges.
If the patch release version or adjacent vulnerable version is not available in the repository, we filtered it out together with the associated CVEs from our database.
Then we extracted the diff methods from pairwise class files between the adjacent vulnerable version and the patch release version and checked whether the methods that were patched exist in these diff methods.
{To obtain more accurate \vulroot, we employ the following strategies to discard or retain patch methods for further augmentation: (1)  
Patch methods that exist in neither version (i.e., the patch release version and the adjacent vulnerable version) will be discarded; (2) Patch methods that exist in both versions are directly considered as \vulroot. (3) Patch methods that only exist in the patch release version are newly added patch methods for the adjacent vulnerable version and will be retained for augmentation.}
{During the process of patch method localization in library release versions, we observed the absence of all patch methods for some CVEs, thus, we excluded these CVEs and obtained {\textbf{362}} libraries with {\textbf{14,775}} versions involved in {\textbf{502}} CVEs.}

\begin{figure}
\centering
  \includegraphics[width=0.5\textwidth]{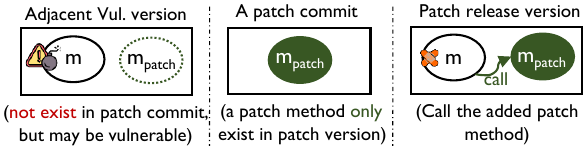}
  \caption{Motivation for Augmentation}
  \label{fig:augment}
\end{figure}

\smallskip
\subsubsection{Augmentation mechanism}
\label{sec:vulroot-aug}
It is common to add a new class or method during vulnerability fixing. Then, the added patch methods are typically called by others, aiming to fix the vulnerability in those methods. 
{Existing work (e.g., Eclipse Steady~\cite{plate2015impact,ponta2018beyond,ponta2020detection}) has overlooked the impact of added patch methods when identifying \vulroot. They argued that these added patch methods are secure and can be ignored. However, our observation reveals that ignoring added patch methods can lead to overlooking vulnerable methods that invoked added patch methods but were absent in patch commits.}
For example, in Figure~\ref{fig:augment}, if the patch method $m_{patch}$ only exists in the patch release version but not in the adjacent vulnerable version, we regarded it as an added patch method. If a method $m$ invoked the added patch method $m_{patch}$ in the patch release version but did not invoke this patch in the adjacent vulnerable version, the method $m$ from the adjacent vulnerable version is still considered vulnerable, even if it did not appear in the patch commit.
{Therefore, such methods should also be augmented as \vulroot.}

\begin{lstlisting}[language=diff,caption=Patch Commit of CVE-2011-2730 \cite{cve-2011-2730_commit},label={lst:exp1},numbers=left]
boolean isJspExpressionActive(PageContext p) {
  ...
  if (sc.getMajorVersion() >= 3) {
-   if (sc.MajorVersion() > 2 || sc.MinorVersion() > 3) {
-    /* Application declares Servlet 2.4+:JSP 2.0 active.
-     * Skip our own expression support.*/
-     return false;
+   if (sc.MajorVersion() == 2 && sc.MinorVersion() < 4) {
+   /* Application declares Servlet 2.3-:JSP 2.0 not active.
+    * Activate our own expression support.*/
+     return true;
  }}
- return true;
+ return false;
}
\end{lstlisting}

\begin{lstlisting}[language=diff,caption=Method Diff between 3.0.5 and 3.0.6 in org.springframework:spring-web,label={lst:exp2},numbers=left]
Object evaluate(Parameters) throws JspException {
  return isExpressionLanguage(attrValue) 
+   && isJspExpressionActive(pageContext) 
    ? doEvaluate(): attrValue;
}
\end{lstlisting}

Considering the situation that methods that invoked the added patch method in the patch release version may be due to the introduction of new functionalities rather than fixing the vulnerability; therefore, our augmentation mechanism is based
\revise{on the following constraint: A method is considered a $VulRoot$ due to augmentation only if the method invoked the added patch method in the patch release version but not in the adjacent vulnerable version.}
In other words, there are no other changes in the augmented $VulRoot$ except for the call relationship to the added patch methods.

For a real case, the TPL ``org.springframework:spring-web''
is affected by the CVE-2011-2730~\cite{cve-2011-2730}, causing multiple versions (the versions before 2.5.6.SEC03, and 3.0.0$\sim$3.0.6) to be vulnerable. 
CVE-2011-2730 is caused by evaluating Expression Language (EL) expressions in tags twice, which allows remote attackers to obtain sensitive information.
As shown in Listing~\ref{lst:exp1}, the developers only activate their expression support when the application declares Servlet 2.3- (Lines 8-11) and set ``springJspExpressionSupport'' to false by default (Line 14), avoiding the potential double EL evaluation problem on pre-Servlet-3.0 containers, which indicates that this method acts as a bug fix.
Although this patch method is shown as modified in the patch commit, however, we found that it only existed in patch versions (2.5.6.SEC03 and 3.0.6).
Therefore, the method ``\texttt{isJspExpressionActive()}'' is an added patch method for vulnerable versions.

To further confirm the impact of the added patch method on fixing the vulnerability, we checked its call relationships in the patch release version (V3.0.6).
We found that five methods directly called this added method and all of them existed in the adjacent vulnerable version (V3.0.5).
For example, in Listing~\ref{lst:exp2}, the method ``\texttt{evaluate()}'' called the added patch method ``\texttt{isJspExpressionActive()}'' (Line 3) in V3.0.6 to fix CVE-2011-2730, and it still existed in V3.0.5 without invoking the added patch method.
\revise{Therefore, this method located in V3.0.5 is vulnerable and should be augmented to the list of vulnerable root methods.}
{Unfortunately, all of the patch commits did not record such call relationship,}
thus existing work only based on patch commits cannot identify the in-depth \vulroot, while \tool augments the \vulroot with such vulnerable methods via multi-version analysis.

\begin{algorithm2e}[t]
    \footnotesize
 \setcounter{AlgoLine}{0}
 \caption{{Vulnerable Root Method Augmentation}}
 \label{alg:1}
 \DontPrintSemicolon
 \SetCommentSty{mycommfont}
 %\SetAlgoLined
 {
     \KwIn{$m_0$: an added patch method. $P_{cg}$: the call graph of patch release version, $V_{cg}$: the call graph of the adjacent vulnerable version. }
     \KwOut{$R$: Vulnerable root methods based on $m_0$.}
     $Visit \gets \varnothing$\;
     $Q \gets Queue()$\;
     $Q.push(m_0)$\; 
     $Visit \gets Visit$ $\cup$ $\{m_0\}$\;
     \While{$Q \neq \varnothing$}{
        $m \gets Q.pop()$\; 
        $S_m \gets$ getCaller($m$, $P_{cg}$) \tcp*[h]{Get direct callers of $m$.}\;
        \If{$S_{m}$ $=$ $\varnothing$}{
            $continue$\;
        }
         \ForEach{$c$ $\in$ $S_m$}{
            % \tcp*[h]{If $c$ exists in $V_{cg}$.}\;
            \If{$isInGraph(c, V_{cg})$}{ 
                $R \gets R \cup \{c\}$  \tcp*[h]{Incorporate it into the results.}\;
            }
            \Else{
                \If{$c$ $\notin$ $Visit$}{
                    $Q.push(c)$\;
                    $Visit \gets Visit \cup \{c\}$\;
                }
            }
        }
        \If{$R \neq \varnothing$}{
            \KwRet{$R$}
        }
    }
}
\end{algorithm2e}

Algorithm~\ref{alg:1} details the augmentation procedure. Given an added patch method $m_0$, the call graph of the adjacent vulnerable version and patch release version ($V_{cg}$ and $P_{cg}$ respectively), \tool outputs the augmented \vulroot $R$ based on $m_0$.
In detail, we leverage the function call relationship of the added patch methods in the patch release version, to mine the methods in the call chain that exist in the adjacent vulnerable version (Lines 5-18).
In particular, for each added patch method $m_0$, if it is invoked by other methods in the patch release version, we will check whether these callers exist in the adjacent vulnerable version (Line 11).
If exists, the caller will be augmented into the set of \vulroot (Line 12), otherwise, it will be added into the queue for further mining \vulroot (Lines 13-15).
Note that, once we obtain the results of \vulroot, we will exit the while loop directly (Lines 17-18), to avoid increasing the negative impact of the possible errors of the added patch methods.
After the above process, the set of augmented \vulroot is constructed.

\subsection{Vulnerable API Identification}
\label{sec:identifyAPI}
Based on the final \vulroot identified in Section \ref{sec:vulroot}, in this section, we aim to mine the \textit{\textbf{\vulapi}} via call graph, which is defined in Section~\ref{sec:key-term}.
We mine all the \vulapi because if a project invokes an API of a library that eventually reaches or calls the vulnerable root method, then this API should also be regarded as vulnerable. In fact, according to our observation, the \vulroot are hardly invoked by projects directly.
Therefore, we also mine and maintain all the \vulapi for each vulnerable library version for further analysis.

Specifically, for each vulnerable {library}, we mined for the \vulapi affected by the \vulroot based on backward call graph analysis.
% Firstly, we generated the call graph of the library by employing the RTA algorithm of Soot~\cite{soot},
{Firstly, we generated the call graph of the library by employing context-insensitive points-to analysis provided by the static framework Tai-e~\cite{tan2022tai}}
and considered all the methods as the entry points to obtain a complete call graph.
Subsequently, starting from the \vulroot, we traversed their called traces in the call graph and recorded all the methods executed in the traces. 
% These recorded methods are \vulapi for the single vulnerable package, which are indirectly threatened by vulnerabilities.
In such a manner, we obtained all the \vulapi for each vulnerable library version.
%Since the call relationship varies between different versions of the same package, we need to perform the above operation once for all package versions within the vulnerable version range affected by the CVE.

\smallskip
\noindent \textbf{Database construction.}
%According to the \vulroot and \vulapi, we constructed a vulnerable method database with the mapping relation: library version to CVEs to vulnerable methods, denoted by \textit{libV-CVE-Vul.M}.
Based on the identified \vulapi, we constructed a vulnerable API database with the mapping relation: library version to CVEs to \vulapi, denoted by \textit{libV-CVE-Vul.API}.
Specifically, we crawled all the vulnerability data and patch commits corresponding to the vulnerability from Snyk Vulnerability DB and GitHub Advisory (as of Feb. 2023) and downloaded the vulnerable {libraries} from Maven~\cite{maven} to support our database.
{Since some versions are not available from Maven or some patch class files do not exist in the {libraries}, we filtered them out.}
% Since some packages or their patch versions are not available for downloading on Maven, we filtered them out.
% We also store the CVEs that affect each vulnerable library version.
We employ the approach above for each CVE in the vulnerable {library}, to obtain a set of \vulapi and construct the vulnerable API database.
% Table~\ref{tab:rq1} shows the detailed information of the constructed database.
Table~\ref{tab:rq1} provides detailed information about the database.
The column ``\#Vul. API (excl. root)'' represents the number of \vulapi obtained from the backward analysis of call graphs. 
``\#Vul. root method'' represents the number of \vulroot, including ones directly obtained from patch commits (``\#Commit'') and the augmented ones (``\#Augm.'') mined by \tool.
We used two counting methods for \vulapi across different library versions: single counting ('API-once') and multiple counting ('API-multi.'). Identical APIs were determined by normalizing their function bodies and comparing hash values.
% 
% \revise{To reflect the data in the vulnerable API database accurately, we conducted two counting methods for the same vulnerable APIs across different library versions: single counting (``API - once''), and multiple counting (``API - mult.''). Specifically, we normalized corresponding function body of each vulnerable API (e.g., remove comments, newlines, and non-ASCII characters, and convert all characters to lowercase), and extracted the hash value for each normalized API. Two vulnerable APIs with the same name and hash value were considered identical. 
The database contains 90,749 unique \vulapi (2,410,779 across library versions) from {362} unique libraries with {14,775} library versions, involved in {502} CVEs.}
%
% Our augmentation mechanism has supplemented 3,359 \vulroot from 42 libraries (with 1,365 versions) and related to 49 CVEs.
\revise{On average, our augmentation mechanism has supplemented 5.9 augmented \vulroot per library and 2.5 per library version, related to 49 CVEs.}
% Note that since methods that are augmented or directly obtained from patch commits may exist in multiple vulnerable versions, such methods are counted multiple times.
% The database is generated by continuous iteration, i.e., if the library is not found in our database during the detection process, we will search for relevant vulnerability information based on the name of the library to continuously extend the database.

\begin{table}
\caption{Statistics of \vulapi in libraries identified by \tool. (LibV.: library versions)}
\label{tab:rq1}
\scalebox{0.95}{
% added vertical line color by fy
% \begin{tabular}{c!{\color{blue}\vline}c!{\color{blue}\vline}c!{\color{blue}\vline}cc}
% \begin{tabular}{c|c|c|cc}
% \hline
% \multirow{2}{*}{\textbf{-}}& \multirow{2}{*}{\textbf{\#Total}} & \multirow{2}{*}{\begin{tabular}[c]{@{}c@{}}\textbf{\#Vul. APIs} \\ \textbf{(excl. root)}\end{tabular}} & \multicolumn{2}{c}{\textbf{\#Vul. root methods}} \\ \cline{4-5} 
%  &  &  & \textbf{\#Commit} & \textbf{\#Augm.} \\ \hline
%  \textbf{Vul. APIs} & \revise{90,749} & \revise{87,417} & \multicolumn{1}{c|}{\revise{3,732}} & \textbf{\revise{249}} \\ \hline
% \textbf{Vul. APIs} & 2,410,779 & 2,348,684 & \multicolumn{1}{c|}{58,736} & \textbf{3,359} \\ \hline
% \textbf{Lib (LibV.)} & 362 (14,775) & 304 (11,619) & \multicolumn{1}{c|}{358 (14,620)} & \textbf{42 (1,365)} \\ \hline
% \textbf{CVEs} & 502 & 405 & \multicolumn{1}{c|}{493} & \textbf{49}  \\ \hline
% \end{tabular}}
\begin{tabular}{cc|c|c|cc}
\hline
\multicolumn{2}{c|}{\multirow{2}{*}{\textbf{-}}} & \multirow{2}{*}{\textbf{\#Total}} & \multirow{2}{*}{\textbf{\begin{tabular}[c]{@{}c@{}}\#Vul. API\\ (excl. root)\end{tabular}}} & \multicolumn{2}{c}{\textbf{\#Vul. root method}} \\ \cline{5-6} 
\multicolumn{2}{c|}{} &  &  & \multicolumn{1}{c|}{\textbf{\#Commit}} & \textbf{\#Augm.} \\ \hline
\multicolumn{1}{c|}{\multirow{2}{*}{\revise{\textbf{API}}}} & \revise{\textbf{once}} & \revise{90,749} & \revise{87,417} & \multicolumn{1}{c|}{\revise{3,732}} & \textbf{\revise{249}} \\ \cline{2-6} 
\multicolumn{1}{c|}{} & \textbf{\revise{mult.}} & 2,410,779 & 2,348,684 & \multicolumn{1}{c|}{58,736} & \textbf{3,359} \\ \hline
\multicolumn{2}{c|}{\textbf{Lib (LibV.)}} & 362 (14,775) & 304 (11,619) & \multicolumn{1}{c|}{358 (14,620)} & \textbf{42 (1,365)} \\ \hline
\multicolumn{2}{c|}{\textbf{CVE}} & 502 & 405 & \multicolumn{1}{c|}{493} & \textbf{49} \\ \hline
\end{tabular}}
\begin{center}
    \footnotesize
    \textit{Note: \textbf{excl. root} - Vulnerable APIs that exclude \vulroot; \textbf{once} - The same vulnerable API is counted once across versions; \textbf{mult.} - The same vulnerable API is counted multiple times across versions.}
\end{center}
\end{table}

\subsection{Used Vulnerable API Detection}

In this section, we describe how to detect whether the \vulapi from TPLs are used in projects.
For a given Java project with its used libraries, 
% we use the same strategy above to generate its call graph based on Soot,
{we generate its call graph by employing the context-insensitive points-to analysis of Tai-e \cite{tan2022tai},}
%RTA algorithm of Soot~\cite{soot},
which is the bedrock to determine whether it invokes \vulapi.
If it depends on a library version in the vulnerable API database, we search out the used \vulapi from the database for this library.
Specifically, for each method in the call graph of the project, we analyze whether it invokes the \vulapi in the library, if true, \tool marks the \vulapi used by developers.
% \tool extracts the vulnerable call path from the entry of the project to the vulnerable root methods. 
Besides, it also reports the vulnerable dependency, the used \vulapi in the library, the call frequency of \vulapi, and the involved CVEs.
Suppose all the methods in the project do not call the \vulapi, in that case, the project uses the vulnerable library without using the vulnerable code, which should not be regarded as vulnerable usage.
%we take the intersection of the methods in the project's call graph and the set of vulnerable methods in the library, to generate a vulnerable method using report, which contains vulnerable dependency, used vulnerable methods in library, call frequency of vulnerable methods and involved CVEs.
%If the intersection is empty, it proves that although the project contains this vulnerable third-party library, however, it does not actually use the \vulapi in the library.

\section{Evaluation}
In this section, we evaluated \tool on real-world projects to answer the following research questions:

\noindent \textbf{RQ1:}
Can \tool effectively identify vulnerable root methods {and \vulapi}?

\noindent \textbf{RQ2:}
Can \tool outperform state-of-the-art tools in {detecting} vulnerable projects threatened by vulnerable third-party libraries?

\noindent \textbf{RQ3:} How do the \revise{sifting} and augmentation mechanisms contribute to vulnerable API detection for \tool?

\noindent \textbf{RQ4:}
How is the status quo of vulnerable libraries used in open-source projects?

\subsection{RQ1: Effectiveness Evaluation}\label{sec:rq1}

\subsubsection{Setup}
Given that \vulapi are derived from the backward call graph analysis, it can be reasonably assumed that APIs directly or indirectly calling the \vulroot may also contain vulnerabilities. As a result, the accuracy of \vulapi depends on the accuracy of \vulroot. 
This experiment aims to investigate the effectiveness of \tool in identifying \vulroot and \vulapi, and take an in-depth analysis of root causes. 
Specifically, this experiment is based on our database containing {362} vulnerable TPLs ({14,775} library versions), involving {502} CVEs.
{Due to the lack of ground truth for \vulroot correlated with CVEs, we manually analyze the \revise{sifted} patch-unrelated methods and augmented \vulroot to assess the effectiveness of \revise{sifting} and augmentation mechanisms. Furthermore, we additionally provide the ground truth for \vulapi to validate the vulnerable API database and also perform an error analysis to estimate the effectiveness of the vulnerable API database provided by \tool.}

\begin{lstlisting}[language=diff,caption=Patch patterns with examples, label={lst:patterns},numbers=left]
// Pattern 1: Checker
+ boolean checkPathSecurity(String path){ 
+   contain_ = path.contains("../");
+   end_ = path.endsWith(".log")
+   if (!StringUtils.isBlank(path)) {
+     if ( start_ && !contain_ && end_ ) {
+       return true; }} 
+   return false; }
// Pattern 2: Filter
+ String filterSensitive(String url){ 
+   String resultUrl = url;
+   if (containsIgnoreCase(url, _SENSITIVE)) {
+     resultUrl = replaceIgnoreCase(url,_SENSITIVE,_FALSE);}
+   return resultUrl; }
// Pattern 3: Configuration
+ boolean isSupportActive(PageContext pc) {
+   ServletContext sc = pc.getServletContext();
+   String EXP_SUPPORT_CONTXT = "springJspExpressionSupport"
+   String Support = sc.getInitParam(EXP_SUPPORT_CONTXT);
+   if (Support != null) {
+     return Boolean.valueOf(Support);}
+   if (sc.getVersion() >= 3) {
+     Int maj_v = sc.getEffectiveMajorVersion()
+     Int min_v = sc.getEffectiveMinorVersion()
+     if (maj_v==2 && min_v<4) {
+       return true;}}
+   return false;}
// Pattern 4: Enhancer
+ String randomString(int byteLength) {
+   byte[] bytes = new byte[byteLength];
+   SECURE_RANDOM.nextBytes(bytes);
+   CharSet sc = StandardCharsets.ISO_8859_1;
+   return new String(bytes, sc);}
// Pattern 5: Assistance
+ ObjectMapper createVaadinConnectObjectMapper(
+   ApplicationContext c) {
+   ObjectMapper objMapper = 
+     Jackson2ObjectMapperBuilder.json().build();
+   JacksonProperties jacksonProperties = 
+     c.getBean(JacksonProperties.class);
+   if (jacksonProperties.getVisibility().isEmpty()) {
+     objMapper.setVisibility(PropertyAccessor.ALL,
+     JsonAutoDetect.Visibility.ANY);}
+   return objtMapper;}
\end{lstlisting}

\subsubsection{Result}\label{sec:rq1_result}

\begin{table}
\small
\caption{Libraries, CVEs affected by \revise{unique} added \vulroot and projects invoking these libraries}
\scalebox{0.9}{
\begin{tabular}{c|c|c|c|c|c}
% \begin{tabular}{c!{\color{blue}\vline}c!{\color{blue}\vline}c!{\color{blue}\vline}c!{\color{blue}\vline}c!{\color{blue}\vline}c}
% \arrayrulecolor{blue}\hline
\hline
\multirow{2}{*}{-} & \multicolumn{5}{c}{\textbf{\#Augmented \vulroot}} \\ \cline{2-6} 
& \multicolumn{1}{c|}{\textbf{0}} & \multicolumn{1}{c|}{\textbf{\revise{1$\sim$5}}} & \multicolumn{1}{c|}{\textbf{\revise{5$\sim$10}}} & %\multicolumn{1}{c|}{\textbf{10$\sim$15}} &
\multicolumn{1}{c|}{\textbf{\revise{10$\sim$20}}} & \multicolumn{1}{c}{\textbf{\textgreater{}=20}} \\ \hline
\textbf{CVE} & 453 & \revise{36} & \revise{6} & \revise{6} & \revise{1} \\ \hline
\textbf{Lib (LibV.)} & 339 (13,925) & \revise{32 (1,566)} & \revise{7 (6)} & \revise{5 (30)} & \revise{1 (0)} \\ \hline
\end{tabular}}
\label{tab:rq1-added}
\end{table}

\revise{Table~\ref{tab:rq1} shows that \tool can identify 90,749 unique \vulapi (2,410,779 with library versions)}.
Details are aforementioned in the dataset construction of Section~\ref{sec:identifyAPI}.
In the following, we aim to demonstrate the validity of the augmented \vulroot, the \revise{sifted} patch-unrelated methods {and \vulapi}.

\smallskip
\noindent\textbf{(1) Result of augmented \vulroot.}
Table \ref{tab:rq1-added} shows the number of libraries (library versions) and CVEs affected by augmented \vulroot.
Columns 3-6 indicate that more \vulroot are mined compared with those only extracted from patch commits.
\revise{Since there is no single library version with more than 20 unique vulnerable root methods augmented, the value of ``libV.'' is set to 0.}
We manually analyzed each method and summarized five patch patterns (\Cref{lst:patterns}).
These patterns highlight the scenarios in which developers address the vulnerabilities by introducing new patch methods.

\noindent \textbf{P$_{1}$: Checker.} 
To fix vulnerabilities reported in CVEs, developers sometimes add check mechanisms (e.g., add logic statements) to check the legitimacy of the input or improve the original check mechanism. 
For example, Listing \ref{lst:patterns} shows an added method ``\texttt{checkPathSecurity(..)}'' in CVE-2022-26884~\cite{cve-2022-26884_commit} that checks whether the parameter ``path'' transferred conforms to security, e.g., whether it contains ``../'' which does not meet security requirements and may lead to security problems.

\noindent \textbf{P$_{2}$: Filter.}
Some added methods aim to filter out unexpected input with specific conditions. In such a pattern, legitimate input will be retained, and illegitimate ones will be discarded.
For example, in Listing~\ref{lst:patterns}, to fix CVE-2022-40955,
developers added a new method ``\texttt{filterSensitive(..)}'' in {the} patch commit~\cite{cve-2022-40955_commit} to filter out invalid and sensitive cases and keep the url meeting security requirements.

\noindent  \textbf{P$_{3}$: Configuration.}
To avoid the vulnerabilities caused by the lack of default configuration or misuse of configuration, developers tend to standardize or improve existing configurations.
{As described in CVE-2011-2730, the spring-framework~\cite{spring-framework} suffered from Expression Language Injection. Developers addressed the potential Double EL Evaluation issue by defaulting the relevant parameter `springJspExpressionSupport` to false in their patch commit~\cite{cve-2011-2730_commit}.}
% For example, Listing \ref{lst:patterns} shows an added method ``\texttt{getHandlerContext(..)}'' in {the patch commit}~\cite{cve-2022-26112_commit}.
% To fix the vulnerability resulting from inappropriate default configuration support, developers create a new method to obtain the context so that the groovy function can be disabled by default.

\noindent \textbf{P$_{4}$: Enhancer.}
Developers usually introduce a series of algorithms and operations to enhance existing programs for security, such as introducing more robust algorithms and safer authentications.
{Listing~\ref{lst:patterns} shows an added method ``\texttt{randomString()}'' identified in the patch commit~\cite{CVE-2021-29480_commit}, which provides a randomly generated default value, enhancing the client-side session encryption secret after the update.}
% Listing \ref{lst:patterns} shows an added method ``\texttt{unmarshal(..)}'' in CVE-2017-15703~\cite{cve-2017-15703_commit}.
% In this method, a safer XML reader is created to avoid XXE attack through JAXB, caused by malicious requests with crafted XML payload.    

\noindent \textbf{P$_{5}$: Assistance.}
Some added methods may not directly fix vulnerabilities, but their relevance can be assessed through correlation analysis of commit messages and methods. For example, the added method ``\texttt{createVaadinConnectObjectMapper(..)}'' in the patch commit~\cite{cve-2020-36319_commit}, shown in Pattern 5 of Listing~\ref{lst:patterns}, creates a custom ObjectMapper to help address the vulnerability.
% Some added methods seem not to directly participate in fixing the vulnerabilities, but the relevance to the repair can be extracted with correlation analysis of commit messages and the methods. {The added method ``\texttt{createVaadinConnectObjectMapper(..)}'' in the patch commit~\cite{cve-2020-36319_commit}, as shown in Pattern 5 of Listing~\ref{lst:patterns}, is responsible for creating a custom ObjectMapper object. This object provides users with a customized interface to facilitate the patching of the vulnerability.}
% Listing~\ref{lst:patterns} shows an added method ``\texttt{initialiseSocket(..)}'' in CVE-2018-11775~\cite{cve-2018-11775} %in \cite{activemq} that offers support for hostname verification.    

\begin{figure}[t]
    \centering
    \subfloat[Augmentation mechanism]{
    \includegraphics[width=0.25\textwidth]{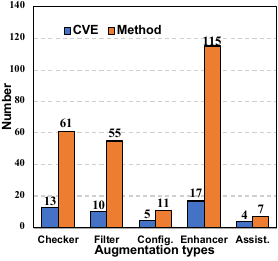}
    \label{Fig:augbar}}    
    %\quad
    \subfloat[\revise{Sifting} mechanism]{
    \includegraphics[width=0.22\textwidth]{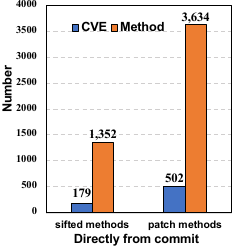}
    \label{Fig:filterbar}
    }  
    \label{fig:augfilter}
    \caption{\revise{Unique augmented methods, \revise{sifted} patch-unrelated methods and corresponding CVEs}}
\end{figure}

For the 5 types of added patch patterns, we further investigated the number of \vulroot that are augmented due to each type as well as the CVEs involved. 
Figure~\ref{Fig:augbar} shows the result.
Among the {49} CVEs supplemented with \vulroot, 
{13} CVEs and {17} CVEs are fixed by adding a \textbf{Checker} and \textbf{Enhancer} in patch commits, respectively, which shows that they are the common fix solutions.
Moreover, we augment \revise{249 unique} methods into \vulroot in total, and \revise{115} methods (the most) are augmented by \textbf{Enhancer}. 
% Note that we only counted the number of methods augmented for corresponding CVEs and did not repeatedly count the same methods included in different vulnerable versions.
{As for the analysis of augmented \vulroot in patch commits, our validation strategy unfolds in two steps: first, we validate whether the added patch method associated with augmented root methods achieves the patching effect; second, we check whether the augmented root method was defective before invoking the added patch method. If the added patch method is patch-unrelated, or if the augmented root method was secure in the adjacent vulnerable version of the library, we determine that this augmented root method was an FP. We validate the above steps for 49 CVEs affected by the augmentation mechanism. Out of the \revise{249} augmented \vulroot, \revise{16} functions (involving 6 CVEs) were confirmed as FPs, achieving \textbf{\revise{93.57\%}} precision.
% Among these, 2 functions were identified as FPs because their associated patch methods were unrelated to the patch, while the remaining 4 were due to augmented root methods being secure, not containing the vulnerability.
}

\smallskip
\noindent\textbf{(2) Result of \revise{sifted} patch-unrelated methods.}
\Cref{Fig:filterbar} displays the number of involved CVEs and \revise{sifted} patch-unrelated methods, including {1,352} \revise{sifted} methods associated with {179} CVEs. Since the \revise{sifting} mechanism involves a large number of methods, we conducted manual analysis on 50 randomly selected CVEs to evaluate the effectiveness (i.e., precision and recall) of the \revise{sifting} mechanism. The sample set consisted of {807} initial patch methods, after manual analysis, {298} methods were identified as patch-unrelated and served as ground truth.
\revise{Note that since we are evaluating the effectiveness of the patch-unrelated sifting mechanism, we consider correctly sifting out patch-unrelated methods as a true positive. Therefore, incorrectly sifting out the patch-related method is considered a false positive, and incorrectly identifying and retaining a patch-unrelated method is considered a false negative.}

Overall, our \revise{sifting} mechanism identifies {258} methods patch-unrelated,
achieving an impressive precision of {\textbf{98.06\%}},
with only {5} methods mistakenly considered as invalid patches.
{The reason is that developers move code snippets from one place to another (e.g., into an if clause), which changes the code semantics and causes false positives of \tool.}
As for the false negatives, {45} patch-unrelated methods were not recognized successfully, resulting in a recall rate of {84.90\%}.
The reasons are as follows:
(1) Certain methods have undergone intricate modifications, limiting the \revise{sifting} mechanism.
(2) Method changes before and after patch commits are semantically equivalent. As \tool employs ASTs to extract the changed code, it cannot recognize semantic equivalence.
{For example, in the patch commit~\cite{cve-2014-0193_commit}, the function ``\texttt{protocolViolation(ChannelHandlerContext, String)}'' was split into two functions, with one calling the other.}
% For example, in {the patch commit}~\cite{cve-2022-24280_commit}, the method ``\texttt{DirectProxyHandler:channelActive(..)}'' only canceled null assignment for the variable ``command'' and deleted the return statement. 
However, \tool fails to recognize it as patch-unrelated, leading to a false negative.

\revise{Furthermore, we employed Wilson's score confidence interval~\cite{agresti1998approximate} to calculate the real false positive rate (FPR) and false negative rate (FNR) of the sifting mechanism, which requires solving for $p$ in the following formula:}

\revise{
\begin{equation}
\label{eq:1}
    \left|p-\hat{p}\right|=z\cdot\sqrt{\hat{p}\cdot\left(1-\hat{p}\right)/n}
\end{equation}
}

\noindent{\revise{where $p$ is the real FPR or FNR, representing the probability of FPs or FNs in the overall population; $\hat{p}$ is the estimated FPR or FNR, representing the proportion of FPs or FNs calculated from the sample $n$; and $z=1.96$ is the critical coefficient for a 95\% confidence interval.
Thus, the FPR of the sifting mechanism is 0.98\% with a 95\% confidence interval (CI) of [0.42\%, 2.28\%], and the FNR is 15.10\% with a 95\% CI of [11.48\%, 19.61\%].}}

\begin{algorithm2e}[t]
    \footnotesize
 \setcounter{AlgoLine}{0}
 \caption{{Construction of Ground Truth for Vulnerable APIs}}
 \label{alg:2}
 \DontPrintSemicolon
 \SetCommentSty{mycommfont}
 %\SetAlgoLined
 {
     \KwIn{$A_{db}$: vulnerable API database, $R_{sam}$: sampled vulnerable root methods, $R_{err}$: the false positives of vulnerable root methods in $R_{sam}$.}
     \KwOut{$A_{sam}$: sampled vulnerable APIs, $A_{err}$: the false positives of vulnerable APIs in $A_{sam}$.}
    $A_{sam}$ $\leftarrow \emptyset$\;
    $A_{err}$ $\leftarrow \emptyset$\;
    \ForEach{$vulAPI \in A_{db}$}{
        \tcp*[h]{Get associated vul. root methods of $vulAPI$.}\;
        $vulRoots \gets \text{getSourceRoots}(vulAPI)$\;
        \If{$R_{sam} \cap vulRoots == \varnothing$}{
            $continue$\;
        }
        $A_{sam} \gets A_{sam} \cup \{vulAPI\}$\; 
        $isErrAPIFlag \gets True$\;
        \ForEach{$vulRoot \in vulRoots$}{
            \If{$vulRoot \notin R_{err}$}{
                $isErrAPIFlag \gets False$\;
                $break$\;
            }
        }
        \If{$isErrAPIFlag$}{
            $A_{err} \gets A_{err} \cup \{vulAPI\}$\;
        }
    }
    \KwRet{$A_{sam}$, $A_{err}$}\;
}
\end{algorithm2e}

\smallskip
\noindent{\textbf{(3) Result of \vulapi.}}
{
% Due to the vast number of \vulapi, constructing a ground truth for \vulapi is necessary to facilitate a sampling analysis.
In light of the need to validate the experiments' results, including the effectiveness of \vulapi (RQ1), the comparison experiment (RQ2), the ablation study (RQ3), and the large-scale analysis (RQ4), we have established a common ground truth for evaluating these experiments. Specifically, given the large number of CVEs associated with the vulnerable APIs in RQ1, the overlapping APIs detected in RQ3, and the detection results in RQ4, it is impractical to analyze each vulnerable API individually. Therefore, we chose to conduct a sampling analysis on them and selected CVEs relevant to the detection results of RQ2 and the non-overlapping vulnerable APIs detected in RQ3. 
\revise{Finally, the \textit{ground truth}'s data sources were from 58 CVEs. 
% Considering that counting vulnerable APIs multiple times across library versions inflates the total number of the sampled vulnerable APIs, thereby affecting the accuracy of vulnerable API evaluation, we employed the single counting method as described in Section~\ref{sec:identifyAPI}.
These 58 CVEs involved 26,720 unique \vulapi, which were directly or transitively reached from 270 unique \vulroot.}
% Finally, we obtained a sample set of 58 CVEs involving 981,805 vulnerable APIs (including 280 vulnerable root methods) to obtain the data sources of this \textit{ground truth}.}
% Given the necessity of validating subsequent experiments, which include evaluating the effectiveness of vulnerable API database (RQ1), the comparison experiment (RQ2), non-overlapping detected vulnerable APIs and overlapping APIs in the ablation study (RQ3), and the large-scale analysis (RQ4), it is essential to establish the ground truth for vulnerable APIs. Because of the large number of vulnerable APIs in the database, validating each one individually is not feasible. Therefore, we selected several vulnerable APIs related to specific CVEs. Specifically,
% considering the validation for the subsequent experiments including the effectiveness of vulnerable APIs (RQ1), the comparison experiment (RQ2), the ablation study (RQ3), and the large-scale analysis (RQ4). 

\revise{The \vulapi are generated based on \vulroot and function call relationships. 
{Since we have utilized the advanced tool for generating call graphs as the foundation of our research, and validating the accuracy of these call graphs is beyond the scope of our research, we assume that the function call relationships are accurate}
% Since the call graph generation is not our technical contribution, we assume the function call relationships are accurate 
and determine the validity of \vulapi based on the validity of \vulroot. Our validation strategy for these large number of vulnerable APIs in the \textit{ground truth} is as follows: (1) we first manually analyze \vulroot based on patch commits and vulnerability descriptions, following the method described in references~\cite{nguyen2016automatic,sun2022verjava}, to obtain the labels of the 270 \vulroot. (2) As shown in \Cref{alg:2}, we automatically extract the associated \vulroot for each vulnerable API. If all of these \vulroot are not vulnerable, we consider that vulnerable API to be a false positive. This process results in obtaining the labels for the 26,720 unique \vulapi as \textit{ground truth}. Ultimately, we identified 386 of 26,720 unique \vulapi (including 25 of 270 unique \vulroot) as false positives, and the vulnerable API database has a false positive proportion of 1.45\% with a 95\% CI of [1.31\%, 1.59\%].}
% {To construct the \textit{ground truth}, we manually analyzed the sample set aforementioned above, following the method described in references~\cite{nguyen2016automatic,sun2022verjava}, to obtain the labels of the 981,805 vulnerable APIs. Our validation strategy for manual analysis is as follows: (1) we identify the vulnerable root methods associated with the vulnerable API; (2) we analyze whether the {vulnerable} root method is flawed based on the patch commit. Based on the commit messages and vulnerability description, if none of the {corresponding vulnerable} root methods exhibits a flaw, then the vulnerable API is deemed not vulnerable. After manual analysis, we identified 2,470 vulnerable APIs (including 26 of 280 vulnerable root methods) as false positives. Ultimately, the \textit{ground truth} contains 979,335 vulnerable APIs, including 254 vulnerable root methods.}

\smallskip
\noindent \fbox{
\parbox{0.95\linewidth}{
\textbf{Answer to RQ1:} {\tool can effectively augment \vulroot which are absent in patch commits {with \revise{93.57}\% precision} and \revise{sift} out patch-unrelated methods with {98.06\%} precision. Eventually, \revise{we construct a database consisting of 90,749 vulnerable APIs (2.4M with library versions) with 1.45\% false positive proportion with a 95\% CI of [1.31\%, 1.59\%] from 362 TPLs.}
}}}

\subsection{RQ2: Comparison with existing work} \label{sec:rq2}
In this section, we demonstrate the effectiveness of \tool by comparing it with the state-of-the-art tool, 
Eclipse Steady~\cite{plate2015impact,ponta2018beyond,ponta2020detection}, which is the only {open source} tool providing a forward reachability analysis at the method level so far. 
% It takes a project as input and initially identifies TPLs directly or transitively dependent on the project using Project KB~\cite{projKB}, and employs either Soot~\cite{soot} or WALA~\cite{wala} for forward reachability analysis.

\subsubsection{Dataset Collection}
We collected Java projects from GitHub with different numbers of stars. In total, we crawled {13,708} real-world projects with stars ranging from 70,000 to 0, among which 6,416 can be successfully compiled (using ``mvn compile''), while others failed to be compiled due to the use of private libraries or some unpassed plugins.
We further filtered projects that did not depend on the vulnerable library versions in our database, and eventually obtained {3,147} real-world potentially vulnerable projects.

{Steady manages its vulnerability data within Project KB, which includes CVE-related information, including vulnerability descriptions, affected libraries, affected library versions, patch library versions, patch links, and more.}
To ensure a fair comparison with Steady,
% Since Steady and \tool may maintain different CVEs and such differences cannot reflect the difference of the internal methodology, to make a fair comparison, 
we selected the CVEs that are both maintained by Steady and \tool as the comparison dataset, i.e., {213} CVEs in total. We obtained the vulnerable libraries versions affected by these CVEs on GitHub Advisory Database~\cite{gitadvisory}, {171} libraries with {6,153} library versions in total, and
% Based on the {171} libraries with vulnerable versions, we 
finally located {1,045} projects which depended on them.
% and used them for a fair comparison with Steady.

\subsubsection{Setup}
Steady supports static analysis and dynamic-based analysis to analyze the vulnerable code reachability, while the dynamic-based methods require JUnit or application-specific tests, which are often unavailable or insufficient in public Maven projects.
% as input for each project. However, there are no publicly available Maven projects associated with test inputs. Even though few projects do provide access to their test code, they either serve as testing frameworks themselves or exhibit limited and incomplete test coverage, which makes it difficult to locate projects with high quality. 
Therefore, we compare \tool with Steady in terms of static analysis.
%
%Steady takes the scanned project source code as input and initially identifies third-party libraries directly or transitively dependent on the project using Project KB. Subsequently, Steady employs Soot~\cite{soot} or WALA~\cite{wala} to facilitate forward reachability analysis.
{Steady takes a project as input and initially identifies TPLs directly or transitively dependent on the project using Project KB~\cite{projKB}, and employs either Soot~\cite{soot} or WALA~\cite{wala} to facilitate static analysis.}
% Steady takes the project source code as input, initially conducting a traditional software component analysis to identify third-party libraries directly or transitively dependent on the project. Subsequently, Steady utilizes the Soot~\cite{soot} or WALA~\cite{wala} static analysis framework to build the call graph for the project, enabling forward vulnerability reachability analysis.
% Since Steady employs either Soot~\cite{soot} or Wala~\cite{wala} to construct call graphs~\cite{steady-page}, 
To eliminate the side-effect caused by different static analysis frameworks {between \tool and Steady}, we choose Soot as the call graph construction framework of Steady and set up the same configuration as we have done with using Soot.
% To conduct a fair comparison for efficiency,
{As for recording the detection time for Steady and \tool, since only the vulnerability reachability analysis part is focused on, we exclude the time spent on identifying vulnerable libraries and directly record the time spent on reachability analysis.}
% we record the time that Steady spends on the static analysis phase, excluding the time it spends on identifying third-party libraries in projects.
% {vulnerability reachability analysis.}
% For \tool, we record the entire time spent in processing projects, including building projects' call graphs and detecting the used \vulapi.

We run Steady and \tool on the aforementioned {1,045} projects,
and compare 
% the efficiency and the number of projects that are identified as vulnerable.
{the effectiveness of detecting vulnerable projects.}
One project identified as vulnerable means that there exists at least one execution path from the project to the vulnerable API of the vulnerable library.

\begin{table}
\small
\caption{Comparison with Steady in terms of the detected cases, involved vulnerable projects, libraries, CVEs, and the time cost.}
\scalebox{0.9}{\begin{tabular}{c|c|c|c|c|c}
\hline
\textbf{-}            & \textbf{\#Cases} & \textbf{\#Projs} & \textbf{\#Libs} & \textbf{\#CVEs} & \textbf{Avg Time (s)} \\ \hline
\textbf{\tool}        & 214              & 177                 & 32              & 42              & 353                     \\ \hline
\textbf{Steady}       & 95              & 66                  & 12              & 13              & 769                     \\ \hline
\textbf{Overlapped}   & 40               & 44                  & 9              & 11              & N.A.                  \\ \hline
\end{tabular}}
\label{tab:rq2_cve}
\end{table}

\subsubsection{Result}
Table~\ref{tab:rq2_cve} shows the comparison results between \tool and Steady. 
The ``Overlapped'' row represents the results identified by both \tool and Steady.
Considering the overall performance, both \tool and Steady can identify vulnerable projects in a finer-grained manner, sharply reducing the vulnerable projects from {1,045} to {177} and {66} respectively. 
Specifically, \tool identified more vulnerable cases than Steady ({214} vs. {95}), {with \tool averaging 353s per project for detection, and Steady averaged 769s.}
% while took much less time (353s vs. 769s).
Besides, {40} cases are both identified by two tools.
{To validate the precision of identified cases scanned by \tool and Steady, we used the \textit{ground truth} for vulnerable APIs proposed in RQ1 to check whether the vulnerable APIs used by projects were false positives.
% we employed the strategy outlined in \Cref{sec:rq1_result} to validate whether the reported \vulapi used in projects were actually vulnerable. 
% We found that the aforementioned 58 CVEs can cover the CVEs involved in \tool and Steady's scanning results. Therefore, we used the 58 CVEs, whose \vulapi have been verified in~\Cref{sec:rq1_result}, as the ground truth. 
% Based on the \textit{ground truth} for vulnerable APIs proposed in RQ1, we checked whether the vulnerable APIs used by projects were false positives.
If there is at least one vulnerable API that is confirmed to be the true positive, the detected case is considered a true positive. Moreover, if one tool identifies a case as a true positive, while another tool does not detect this case, then this case is considered a false negative for the latter. Consequently, we identified 166 (61.71\%) FNs in the scanning results of Steady, while \tool yielded 8 (2.97\%) FPs and 55 (20.45\%) FNs.}
We thus further take an in-depth analysis to investigate the reasons and insights, which are summarized as follows.

\smallskip
\noindent $\bullet$ \textbf{Identified by both tools (40 cases)}.
For vulnerable projects identified by both tools, {these detected cases are all true positives. Furthermore,}
we found that these projects all directly invoked vulnerable libraries, i.e., directly invoked the \vulapi or other APIs of the library which finally reached the \vulroot via call graph. Besides, the \vulroot of these used libraries were all extracted from patch commits, and this is the simplest case that existing patch commit analysis focused on. Therefore, Steady and \tool both can identify them.

\smallskip
\noindent $\bullet$ \textbf{Only identified by Steady (55 cases)}. For projects that were only identified as vulnerable by Steady,
% {these cases are all true positives, and}
% we found that 
some projects invoked vulnerable libraries indirectly.
Since Steady started analysis from the project and further analyzed the direct- and transitive-invoked libraries to detect whether the project became vulnerable through the dependencies, it thus can identify such cases. 
While \tool focused on distilling the vulnerable APIs of each vulnerable library, i.e., only considering the vulnerable libraries directly depend on projects, it cannot identify whether the project can reach vulnerable APIs/code from such transitive dependency.

\smallskip
\noindent $\bullet$ \textbf{Only identified by \tool (174 cases)}.
For the projects only identified by \tool, we found these projects invoked vulnerable library APIs that are not marked as vulnerable by Steady.
There are four possible reasons: (1) {Due to the missing information of vulnerable libraries affected by the same CVE,} Steady exhibits false negatives in the identification of vulnerable libraries, where such vulnerable libraries are mistakenly classified as safe.
For example, the library ``dom4j-2.0.0''
% ~\cite{dom4j}
is suffered by CVE-2020-10683, but Steady fails to identify it as a vulnerable TPL.
% {We checked the vulnerability data~\cite{projKB} maintained by Steady and discovered that the library version ``dom4j-2.0.0'' was indeed in the CVE-2020-10683's vulnerability message provided by Steady, thus the exact reason for this occurrence remains unclear to us.}
(2) Steady identified vulnerabilities based on all the modified and deleted methods in the patch commits. However, if the patch commit added a patch method that is not directly invoked in any other patch commits but is later invoked by methods in the vulnerable library version in another commit,
Steady may not recognize it as vulnerable. In contrast, \tool can mark it as vulnerable owing to its augmentation mechanism.
% For example, the project ``HolgerHees/cloudsync'' directly depends on the library ``bcprov-jdk15on-1.54'' which is affected by CVE-2016-1000346~\cite{cve-2016-1000346}. \tool reported that it invoked the vulnerable API from ``bcprov-jdk15on-1.54'', however, Steady showed that it did not reach the vulnerable code related to CVE-2016-1000346. 
{For example, the project ``jbufu/openid4java'' directly depends on the library ``xercesImpl-2.8.1'' which is affected by CVE-2012-0881. \tool reported that it invoked the vulnerable API from ``xercesImpl-2.8.1'', however, Steady showed that it did not reach the vulnerable code related to CVE-2012-0881.}
After our investigation, we found it indirectly invoked the vulnerable root method augmented by \tool.
Since Steady only extracts the diff methods from patch commits as vulnerable methods, it cannot cope with such a situation, resulting in false negatives.
(3) {When the libraries contain both vulnerable structures and patch structures, Steady is uncertain about whether they include vulnerable code,}
% Steady is uncertain about whether some libraries (actually vulnerable) include vulnerable code, 
resulting in missing some identified results. Steady stored the AST associated with vulnerability to determine whether the current library version contains vulnerable code. Due to some internal errors, the version that is vulnerable is not recognized by Steady.
{(4) The depth of call analysis in forward vulnerability reachability analysis is shallow compared to backward call graph analysis. Forward reachability analysis traces paths from external code to the vulnerability point, emphasizing breadth, while backward call graph analysis starts from the vulnerability point and traces its calling paths outward, focusing more on depth. Consequently, forward reachability analysis lacks the comprehensiveness of backward analysis, as achieving the same depth would require a significant resource investment.}

{As for 8 false positives generated by \tool, they involved 4 CVEs and 4 libraries. The misidentification of these cases stems from the fact that the root methods associated with reported APIs are unrelated to the vulnerabilities. Since the patch involves the addition of member variables with the result of necessitating complex modifications in the initial methods, \tool erroneously determined these root methods were vulnerable before patching.}

\smallskip 
\noindent \fbox{
\parbox{0.95\linewidth}{
\textbf{Answer to RQ2:} 
%\tool can effectively and efficiently detect \vulapi used by projects which outperforms state-of-the-art method-level vulnerability detection tool, Eclipse Steady.
\revise{\tool can enhance the current tool chains by detecting security threats more effectively through deep call chains at the price of potentially missing some cases due to transitive dependencies.}
% outperforms the state-of-the-art method-level vulnerability detection tool Eclipse Steady in analyzing direct dependencies.
}}

\begin{table}
\small
\caption{Ablation study results on different mechanisms. (\Checkmark: Enabled; \XSolidBrush: Disabled; prop.: proportion)}
\scalebox{0.88}{\begin{tabular}{cc|c|c|c}
\hline
\multicolumn{2}{c|}{\textbf{Mechanisms}} & \multirow{2}{*}{\textbf{\#Vul. APIs}} & \multirow{2}{*}{\textbf{\revise{FP prop.(\%)}}} & \multirow{2}{*}{\textbf{\revise{FN prop.(\%)}}} \\ \cline{1-2}
\multicolumn{1}{c|}{\textbf{\revise{Sifting}}} & \textbf{Augm.} &  &  &  \\ \hline
\multicolumn{1}{c|}{\XSolidBrush} & \XSolidBrush & $1,229$ & \revise{$11.89\%\sim16.52\%$} & $2.16\%$ \\ \hline
\multicolumn{1}{c|}{\Checkmark} & \XSolidBrush & $1,158$ & \revise{$6.11\%\sim10.74\%$} & $2.16\%$ \\ \hline
\multicolumn{1}{c|}{\Checkmark} & \Checkmark & $1,183$ & \revise{$6.11\%\sim10.74\%$} & $0$ \\ \hline
\end{tabular}}
% \begin{center}
%     \footnotesize
%     \textit{Note: \textbf{prop.} - proportion}
% \end{center}
\label{rq3:ablation}
\end{table}

\subsection{RQ3: Ablation Study on different mechanisms}
To showcase the contribution of the proposed \revise{sifting} and augmentation mechanisms, we set up an ablation study on them. {Specifically, we execute \tool and \tool- with different mechanisms enabled on the same projects respectively, shown in Table \ref{rq3:ablation}.
The contribution of the augmentation mechanism is not separately studied because it is based on the \revise{sifting} mechanism.}
We then compare the results of the individual scans against each other.

\subsubsection{Dataset Collection}
Our proposed \revise{sifting} and augmentation mechanisms affected {179} and {49} CVEs, respectively, involving {183} libraries with {6,529} library versions. To evaluate the impact of these two mechanisms, 
we selected {1,191} projects that are dependent on the {183} libraries from the {3,147} potential vulnerable projects mentioned in Section~\ref{sec:rq2}, which enables us to assess the contribution of these mechanisms.
% and obtained {1,128} projects in total.

\subsubsection{Result}
\label{sec:rq3_result}
{After scanning these potentially vulnerable projects, \tool identified {284} projects that utilized \vulapi. However, \tool- without any mechanisms, and with only the \revise{sifting} mechanism, detected {293} and {272} projects calling \vulapi, respectively.}
Table~\ref{rq3:ablation} shows the {vulnerable API} detection result of the ablation study. The {``\#Vul. APIs''} column displays the number of detected \vulapi.
We assessed the accuracy of the detected APIs by analyzing the precision of the corresponding \vulroot. \tool decreased {71 (5.78\%)} false positives by employing the \revise{sifting} mechanism, and {25 (2.16\%)} false negatives by utilizing the augmentation mechanism.
{Besides, \tool and \tool- (both without any and augmentation mechanism) identified 1,158 overlapping APIs, \revise{with an 8.13\% false positive proportion with a 95\% CI of [6.11\%, 10.74\%]}.}
Next, we first provide detailed explanations for how \tool achieves the reduction in FPs and FNs through these two mechanisms,
{and subsequently validate the overlapping detected APIs.}

\smallskip
\noindent $\bullet$ \textbf{FP reduction analysis.}
Since \tool- without any mechanisms identified the diff methods before and after the patch commits as patch methods directly, its vulnerable API database may include many non-\vulapi. However, our proposed \revise{sifting} mechanism can \revise{sift} out patch-unrelated methods with high precision, reducing the generation of some non-\vulapi for both \tool and \tool- with the \revise{sifting} mechanism. Therefore, the \revise{sifting} mechanism can eliminate {71} (5.78\%) FPs detected by \tool- without any mechanisms.
% Illustrating the project ``Rogiel/torrent4j'' as a case, it depended on the TPL ``netty-all-4.0.9.Final'' influenced by CVE-2015-2156. \tool- without any mechanisms shows that this project invoked 113 \vulapi which indirectly called the vulnerable root method ``\texttt{DefaultCookie:getMaxAge()}''. However, through meticulous manual verification, we found that it was not a patch method in {the patch commit}~\cite{cve-2015-2156_commit}. In other words, it was not the vulnerable root method in ``netty-all-4.0.9.Final'', causing FPs of \tool- (without any mechanisms).
For example, the project ``gavincook/githubOfflineInstaller'' depended on the TPL ``dom4j-2.0.0-RC1'' influenced by CVE-2020-10683. \tool- without any mechanisms shows that it invoked 3 \vulapi which indirectly called the root method ``\texttt{SAXReader:configureReader(XMLReader,DefaultHandler)}''. However, through meticulous manual verification, we found that it was not vulnerable in {the patch commit}~\cite{cve-2020-10683_commit}, causing FPs of \tool- (without any mechanisms).

\smallskip
\noindent $\bullet$ \textbf{FN reduction analysis.}
Table~\ref{rq3:ablation} reveals that \tool detected {25} additional \vulapi compared to \tool- without augmentation mechanism, indicating that augmentation mechanism can eliminate {25} (2.16\%) FNs. The augmentation mechanism enables \tool to generate more accurate \vulapi. 
For example, the project ``fabric8io/shootout-docker-maven'' utilized the TPL ``tomcat-embed-core-7.0.91'' affected by CVE-2021-30640. In {the patch commit}~\cite{cve-2021-30640_commit}, developers introduced a patch method named ``\texttt{JNDIRealm:doAttributeValueEscaping(String)}'' to implement the necessary escaping. Through our augmentation mechanism, two methods invoking this newly added patch method in the patch release version (V7.0.109) were absent in any other patch commits. This absence resulted in \tool- failing to identify \vulapi related to these augmented \vulroot, leading to FNs in scanning projects.

\smallskip
\noindent $\bullet$ {\textbf{Validation for overlapping APIs.}}
{We used the \textit{ground truth} to validate the overlapped vulnerable APIs detected by both tools. Among the 58 CVEs in \textit{ground truth}, 36 were involved in the ablation study, covering 541 vulnerable APIs out of 1,158 overlapping APIs. We found that 44 out of 541 APIs from \textit{ground truth} were false positives, \revise{and then performed an error analysis using Wilson’s score confidence interval~\cite{agresti1998approximate} to estimate the false positive proportion. Thus, the detected overlapping APIs have an 8.13\% false positive proportion, with a 95\% CI ranging from 6.11\% to 10.74\%.}
}

\smallskip 
\noindent \fbox{
\parbox{0.95\linewidth}{
\textbf{Answer to RQ3:}
\tool effectively reduces FPs by {5.78\%} through \revise{sifting} mechanism and FNs by {2.16\%} through augmentation mechanism, leading to more accurate and comprehensive vulnerable API detection.}
}

\subsection{RQ4: Large-scale Analysis}

Based on the {3,147} projects mentioned in Section~\ref{sec:rq2}, we further conducted a large-scale study 
%on GitHub projects 
by leveraging \tool, to reveal the fact of using \revise{potentially} \vulapi from the vulnerable libraries in real-world projects.

\subsubsection{Impact analysis of \revise{potentially} \vulapi}
\label{sec:rq4_impact_analysis}
Based on the collected dataset, we aim to investigate the impact of \revise{potentially} \vulapi on real-world projects.
The results are shown in Table~\ref{tab:rq4_proj}.
We found that {1,753} projects did not use any of the modules in the vulnerable libraries in our database, {717} projects only used the non-vulnerable modules in the vulnerable libraries, {and 677 projects were \revise{potentially} affected by vulnerable libraries.}
Moreover, we used the \textit{ground truth} for \vulapi proposed by RQ1, to validate the scanning results for conducting a sampling analysis.
These CVEs involve 35 libraries, 134 library versions, and 219 projects using vulnerable modules. Among these 219 projects, TP=215 and FP=4.
\revise{Furthermore, we conducted an error analysis using Wilson’s score confidence interval and found that approximately 21.51\% of all projects have utilized potentially vulnerable modules in the vulnerable libraries. The false positive proportion is 1.83\% with a 95\% CI of [0.71\%, 4.61\%].}
This means that for most projects, even if calling the vulnerable TPL, they are still not affected by the vulnerable library.
{For example, the project ``elibom/jogger'' directly relies on two vulnerable dependencies: jetty-server-8.1.15 and httpclient-4.5.2, and it invoked 9 APIs from jetty-server-8.1.15,}
but none of these APIs were deemed vulnerable.
Thus, it can suspend the processing of these three vulnerable libraries.
Our analysis indicates that vulnerable TPLs may not have a substantial impact on most projects.
We explore the reasons from the following points:
(1) For the vulnerability itself, the vast majority of vulnerabilities threaten only one or specific modules of the software. 
We attempt to maximize the impact range of vulnerabilities in the TPL through backward call graph analysis, to ensure that all the modules \revise{potentially} affected by vulnerabilities are identified.
(2) For the project itself,
it often uses only specific modules from a TPL, not the entire library, meaning it may not invoke potentially \vulapi and thus avoid certain vulnerabilities. 
In large projects that rely on multiple vulnerable libraries, it is crucial to identify if any vulnerable modules are used. This can help developers plan patches and prioritize vulnerability mitigation.

\begin{table}
\small
\caption{Overall status of the real-world projects invoking vulnerable libraries and \revise{potentially} \vulapi.}
\label{tab:rq4_proj}
\scalebox{0.78}{\begin{tabular}{c|c|c|cccc}
\hline
\multirow{2}{*}{} & \multirow{2}{*}{\begin{tabular}[c]{@{}c@{}}\textbf{Not calling} \\ \textbf{vul libs}\end{tabular}} & \multirow{2}{*}{\begin{tabular}[c]{@{}c@{}}\textbf{Calling libs}\\ \textbf{but not vul APIs}\end{tabular}} & \multicolumn{4}{c}{\textbf{Calling libs and vul APIs}} \\ \cline{4-7} 
 &  &  & \multicolumn{1}{c|}{\textbf{1 lib}} & \multicolumn{1}{c|}{\textbf{2 libs}} & \multicolumn{1}{c|}{\textbf{3 libs}} & \multicolumn{1}{c}{\textbf{4+ libs}}  \\ \hline
\textbf{\#Proj} & 1,753 & 717 & \multicolumn{1}{c|}{596} & \multicolumn{1}{c|}{69} & \multicolumn{1}{c|}{11} & \multicolumn{1}{c}{1}  \\ \hline
\textbf{\#CVEs} & N.A. & N.A. & \multicolumn{1}{c|}{73} & \multicolumn{1}{c|}{47} & \multicolumn{1}{c|}{22} & \multicolumn{1}{c}{6}  \\ \hline
\end{tabular}}
\end{table}

\begin{table*}
\caption{Top 5 vulnerable libraries and \revise{potentially \vulapi} being invoked by projects in the dataset.}
\label{tab:rq4_mthd}
\scalebox{1}{\begin{tabular}{c|c|c|l}
\hline
\textbf{ID} & \textbf{Library and version} & \textbf{Frequency} & \textbf{Top invoked \revise{potentially} \vulapi (Frequency)} \\ \hline
\multirow{2}{*}{\textbf{1}} & \multirow{2}{*}{com.alibaba:fastjson:1.2.47} & \multirow{2}{*}{44} & \multirow{2}{*}{\begin{tabular}[c]{@{}l@{}}1. \texttt{JSON:toString()} (171)\\ 2. \texttt{JSON:toJSONString(Object)} (165)\end{tabular}} \\
 &  &  &  \\ \hline
\multirow{2}{*}{\textbf{2}} & \multirow{2}{*}{org.apache.httpcomponents:httpclient:4.5.2} & \multirow{2}{*}{36} & \multirow{2}{*}{\begin{tabular}[c]{@{}l@{}}1. \texttt{CloseableHttpClient:execute(HttpUriRequest)} (72)\\ 2. \texttt{CloseableHttpClient:execute(HttpUriRequest,HttpContext)} (49)\end{tabular}} \\
 &  &  &  \\ \hline
\multirow{2}{*}{\textbf{3}} & \multirow{2}{*}{org.apache.httpcomponents:httpclient:4.5.3} & \multirow{2}{*}{26} & \multirow{2}{*}{\begin{tabular}[c]{@{}l@{}}1. \texttt{CloseableHttpClient:execute(HttpUriRequest)} (61)\\ 2. \texttt{CloseableHttpClient:execute(HttpUriRequest,ResponseHandler)} (10)\end{tabular}} \\
 &  &  &  \\ \hline
\multirow{2}{*}{\textbf{4}} & \multirow{2}{*}{com.alibaba:fastjson:1.2.62} & \multirow{2}{*}{23} & \multirow{2}{*}{\begin{tabular}[c]{@{}l@{}}1. \texttt{JSON:toJSONString(Object)} (63)\\ 2. \texttt{JSONPObject:toString()} (52)\end{tabular}} \\
 &  &  &  \\ \hline
\multirow{2}{*}{\textbf{5}} & \multirow{2}{*}{org.apache.activemq:activemq-all:5.13.2} & \multirow{2}{*}{21} & \multirow{2}{*}{\begin{tabular}[c]{@{}l@{}}1. \texttt{ClassPathXmlApplicationContext:<init>(String)} (27)\\ 2. \texttt{AbstractApplicationContext:getBean(String)} (26)\end{tabular}} \\
 &  &  &  \\ \hline
\end{tabular}}
\end{table*}

\subsubsection{Top vulnerable libraries and \revise{\vulapi}}
We further investigate the most frequently vulnerable libraries and \revise{potentially} \vulapi invoked by projects based on the collected dataset.
Table~\ref{tab:rq4_mthd} shows the result.
The library ``com.alibaba:fastjson:1.2.47''
, a JSON processor, tops with the list with a maximum frequency of {170} invocations of \vulapi.
This is primarily due to the widespread usage of ``\texttt{JSON:toString()}'', which serves as a fundamental functional component of the library.
As TPLs such as ``com.alibaba:fastjson'' are commonly used by numerous developers, the impact of vulnerabilities in TPLs can be highly unpredictable.
Furthermore, as the frequency of calling \revise{potentially} \vulapi increases, the risks within projects escalate accordingly.
Take the project ``luanqiu/java8\_demo''
as an example.
This project directly relies on ``com.alibaba:fastjson:1.2.47'' affected by CVE-2022-25845 and has invoked the \revise{potentially} vulnerable API ``\texttt{JSON:toJSONString(Object)}'' 45 times, indicating that resolving this vulnerable TPL is crucial to mitigate its impact.
This example highlights the importance of promptly addressing vulnerability risks in TPLs when fundamental functional APIs are \revise{potentially} vulnerable.

\smallskip 
\noindent \fbox{
\parbox{0.95\linewidth}{
\textbf{Answer to RQ4:} By leveraging \tool, \revise{we found that only 21.51\% of projects (with 1.83\% false positive proportion and a 95\% CI of [0.71\%, 4.61\%]) were potentially affected by vulnerable TPLs,} which indicates that most coarse-grained detection tools produce many false positives, highlighting the need for more precise analysis.
}}

\section{Threats to validity}
The threats to our work come from the following aspects:
(1) Possible bias of project dataset selection. Since we crawled projects in GitHub according to star numbers, there may be some project deviations. To alleviate it, we tried our best to crawl a large number of real-world projects whose star numbers range from about 70,000 to 0, to make the experiments more representative.
(2) Possible inaccuracy of vulnerable versions of libraries.
There may be inaccuracies in the vulnerable version ranges provided by Snyk Vulnerability DB and GitHub Advisory Database, based on NVD. This can lead to mistakenly identifying a safe version as vulnerable~\cite{dashevskyi2018screening}. To address this, we determined \vulroot by examining adjacent vulnerable versions for patch commits. If \vulroot is not found in earlier vulnerable versions, it indicates that the version is not actually affected, thus minimizing threats and ensuring the validity of our results.
(3) Not consider other semantically equivalent refactoring in the \revise{sifting} mechanism. Since we implement the \revise{sifting} mechanism based on the AST, which focuses on syntax and structure, it cannot comprehensively capture the context of the code. We will consider detecting all semantically equivalent refactoring in our future work.
(4) Possible bias of the ground truth acquisition strategy.
We avoided dynamic testing due to its complex setup and high costs, especially for large codebases. Although vulnerabilities were demonstrated in previous work~\cite{vul4j2022}, the provided repositories didn’t fully support our validation needs for ablation and comparison experiments. Instead, we used a manual validation approach similar to VERJava~\cite{sun2022verjava} and Nguyen et al.~\cite{nguyen2016automatic}.
Besides, since we assume the call graph generation is accurate and determine the validity of \vulapi based on the validity of \vulroot, this strategy may affect the validity of our results.
\revise{(5) Limitation of the static analyzer. Although we used the state-of-the-art call graph generation tool Tai-e, it still has certain limitations because Tai-e~\cite{tan2022tai} is a static analysis framework, which inherently struggles to accurately handle dynamic features, polymorphism, and runtime dependencies, which prevents Tai-e from generating completely precise call graphs.}
\revise{(6) Possible bias arising from different vulnerability data utilized by \tool and Eclipse Steady. Steady manages its vulnerability data within Project KB~\cite{projKB}, which does not completely match the data we collected. This discrepancy may introduce bias in the comparison experiment results.}
\revise{(7) The \vulroot we have augmented are not always vulnerable, which may affect the accuracy of \vulroot.}

\section{Related work}

The most related work to our paper is software composition analysis (SCA)~\cite{SCA_78} of Java projects. 
Plate et al.~\cite{plate2015impact} proposed a {dynamic} analysis to determine if the project could reach vulnerable methods in TPLs.
It was implemented by the {dynamic and static} instrumentation techniques for unit tests and integration tests, respectively.
Ponta et al.~\cite{ponta2018beyond, ponta2020detection} advanced this approach and presented a code-centric and usage-based tool, named Eclipse Steady, to identify the reachability of vulnerable methods or code. Specifically, they first conducted a dynamic analysis to assess the reachability of vulnerable constructs. Then, they used the set of constructs that have actually been executed as the starting point for static analysis. Combining dynamic and static analysis, they found all constructs potentially reachable for vulnerability analysis.
Despite the progress, their dynamic analysis {required} unit tests or integration tests as the input for vulnerability analysis, which {limited} its scalability and effectiveness due to the availability and quality of test code.
\revise{Wang et.al~\cite{wang2020empirical} proposed a bug-driven alerting system that focuses on
security bugs.
In their approach, they directly considered the methods modified in patches as buggy library methods.}
{INSIGHT~\cite{xu2022insight} explores the cross-ecosystem impact of vulnerabilities, specifically determining whether a Python or Java project utilizes a vulnerable C library based on the forward cross-language vulnerability reachability analysis.
}
{Wu et.al~\cite{wu2023understanding} conducted an empirical study aiming to explore the impact of vulnerabilities in upstream libraries on downstream projects. They considered all modified functions in the vulnerability patch as vulnerable functions in libraries. By constructing call graphs for downstream projects and upstream vulnerable libraries, they investigated whether there exists paths in the projects that can invoke the vulnerable functions from the libraries.}
Relying on dependency management tools such as Apache Maven and Apache Ivy, Pashchenko et al.~\cite{pashchenko2018vulnerable,pashchenko2020vuln4real} identified dependencies with known vulnerabilities. They built the paths from projects to their vulnerable dependencies, to address the over-inflation problem when reporting vulnerable dependencies.
In addition, both commercial SCA services (e.g., Snyk~\cite{snyk}, SourceClear~\cite{sourceclear}) and open-source SCA tools (e.g., GitHub Dependabot~\cite{dependabot}, OWASP Dependency Check~\cite{owasp}) {detected} vulnerable TPLs based on vulnerability information from NVD~\cite{nvd}.
Although some SCA commercial tools (e.g., SourceClear~\cite{sourceclear}, and BlackDuck~\cite{blackduck}) support vulnerability reachability analysis, they do not provide open source alternatives, posing a hindrance to executing them. Moreover, their methodology for vulnerability reachability analysis
like Steady's, uses call graph analysis to check if the project invokes \vulapi. Therefore, we only compared \tool with Steady.

There are also other researches that focused on SCA of Android apps, usually known as TPL identification \cite{LibD2017ICSE,libdetect2020ASE,LibID2019issta,libpecker2018,LibRadar2016ICSE,libscout2016ccs,ORLIS2018MOBILESoft,duan2017identifying,chen2015finding,Lili2016SANER,zhan2021atvhunter}.  
Most of them focused on identifying the libraries or library versions used by Android apps via similarity-based or clustering-based methods. Some studies investigated vulnerable TPLs used by projects by detecting whether the projects contained vulnerable TPLs or vulnerable TPL versions~\cite{duan2017identifying,yasumatsu2019understanding,zhan2021atvhunter}. Specifically, OSSPolice \cite{duan2017identifying} maintained a feature database of TPLs, 
and utilized a similarity-based method to identify whether the used library version was vulnerable by comparing it with the vulnerable libraries affected by CVE. {Yasumatsu et al.~\cite{yasumatsu2019understanding} conducted a similar work by using LibScout~\cite{libscout2016ccs} to extract the library versions used by APK and comparing them with vulnerable versions.}
Based on TPLs' feature generation and vulnerability collection, Zhan et al. \cite{zhan2021atvhunter} built a vulnerable TPL database to identify the vulnerable TPL versions used by Android apps.
These studies identified vulnerable TPLs but did not analyze whether the apps accessed the vulnerable code.
In summary, these studies would cause false positives through analysis only at the library level.

As for \tool, we maintain all vulnerable {APIs} for each vulnerable TPL version.
Once projects used a specific library version, \tool can effectively
determine whether the used library version could threaten the projects by analyzing if the projects used \vulapi.

\section{Conclusion}
In this paper, we proposed \tool, a vulnerable API detection system for TPLs, which can precisely find \vulapi used by Java projects. {\tool can \revise{sift} out patch-unrelated methods with high precision, and augment \vulroot which are absent in patch commits,
to identify relatively precise and complete \vulroot.}
Evaluation results show that \tool can effectively
detect \vulapi based on the constructed vulnerable API database and can find the \vulapi and real impact on real projects.
\section*{Acknowledgement}
We thank the reviewers for their insightful comments.
This work was supported by the National Natural Science Foundation of China (No. 62102197, and 62202245), and the Natural Science Foundation of Tianjin (No. 22JCYBJC01010).

\normalem
\bibliographystyle{IEEEtran}
\bibliography{main}

\begin{IEEEbiography}[{\includegraphics[width=1in,height=1.25in,clip,keepaspectratio]{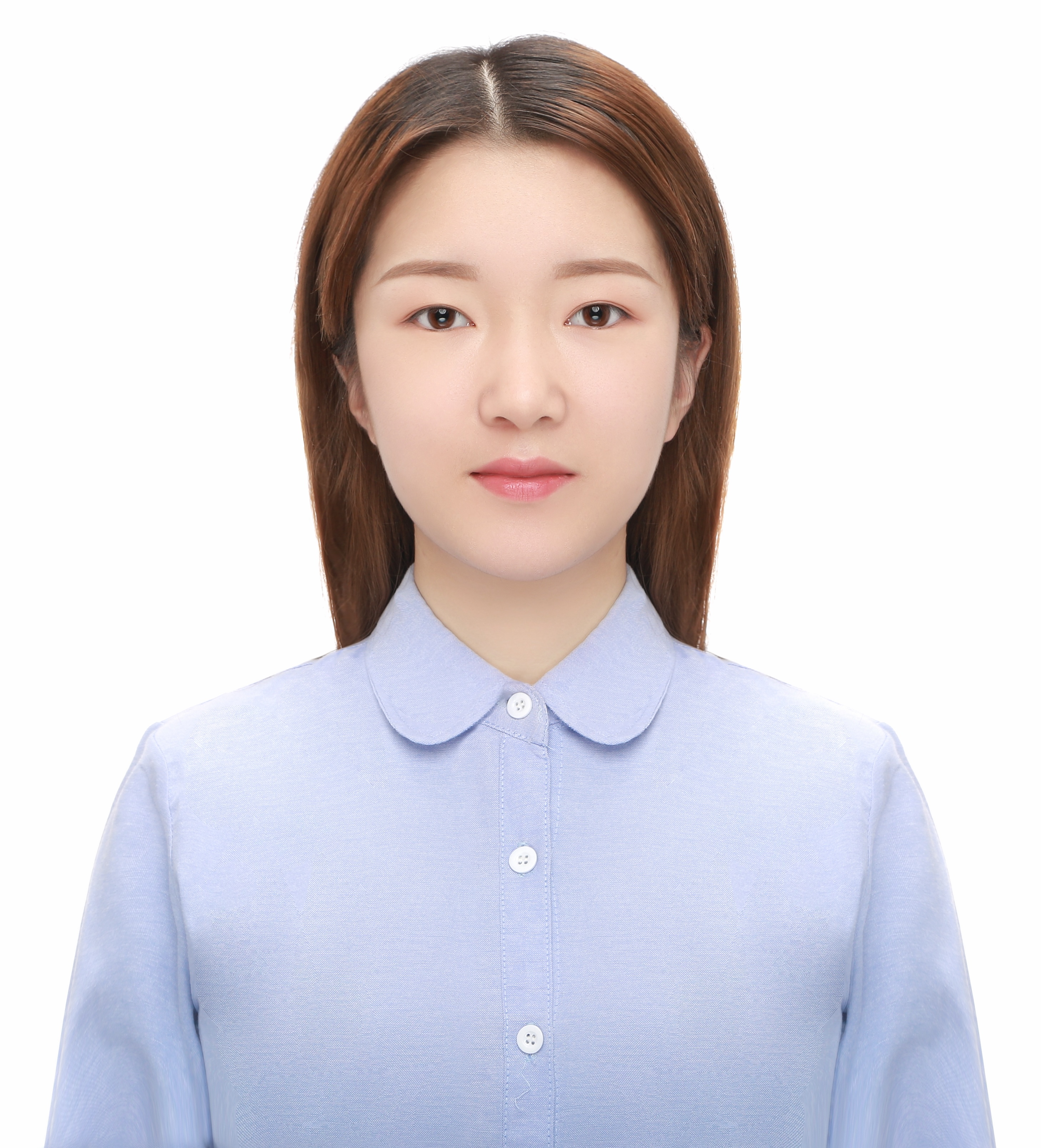}}]
{Fangyuan Zhang} is currently a Ph.D. candidate in the College of Computer Science at Nankai University (NKU). She received her BSc degree in computer science from Jilin University in 2021. Her research focuses on software supply chain security.
\end{IEEEbiography}

\begin{IEEEbiography}[{\includegraphics[width=1in,height=1.25in,clip,keepaspectratio]{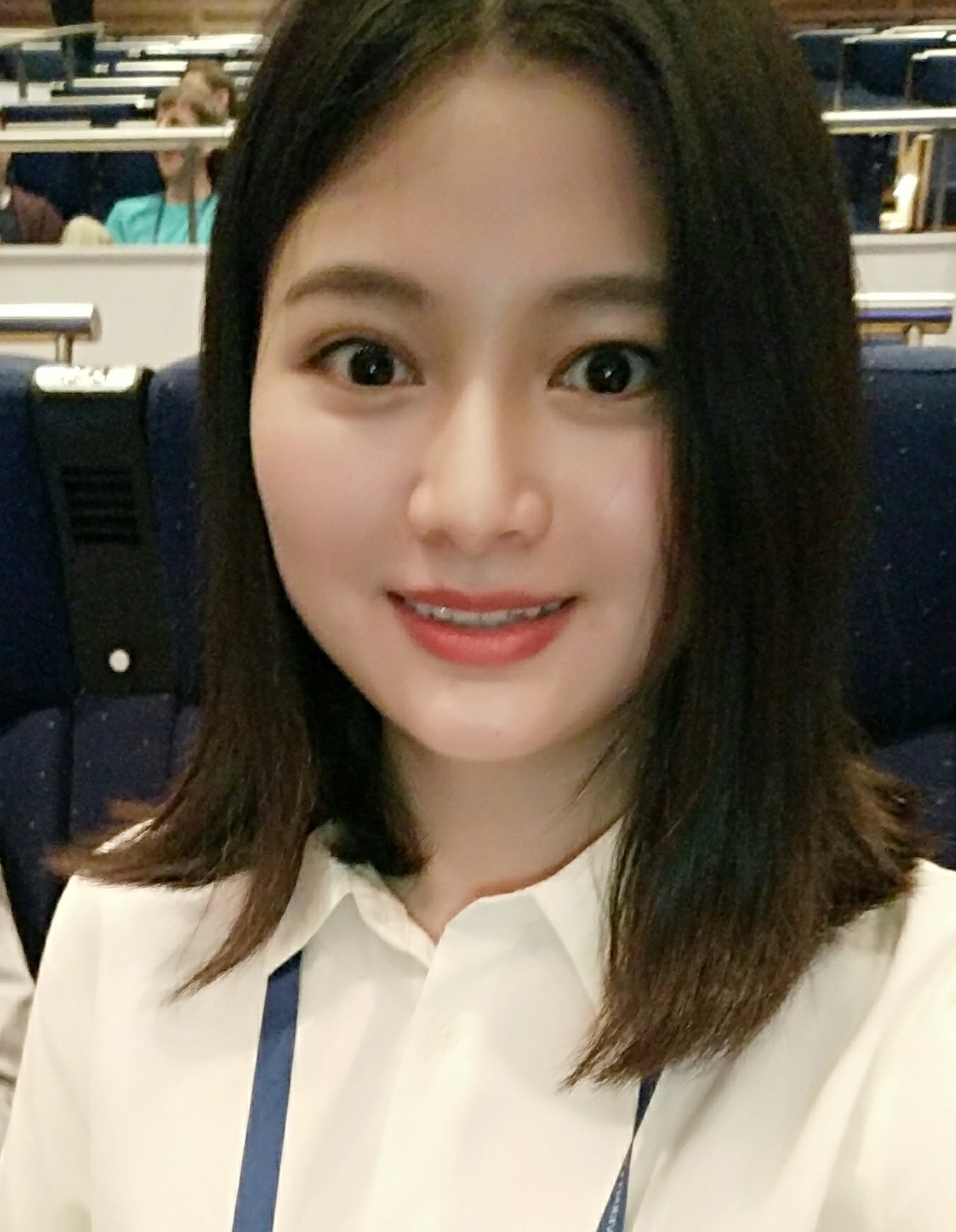}}]
{Lingling Fan} is an Associate Professor at the College of Cyber Science, Nankai University, China. 
In 2017, she joined Nanyang Technological University (NTU), Singapore as a Research Assistant and then had been a Research Fellow of NTU since 2019. Her research focuses on program analysis and testing, and software security. She got 4 ACM SIGSOFT Distinguished Paper Awards at ICSE 2018, ICSE 2021, ASE 2022, ICSE 2023.
\end{IEEEbiography}

\begin{IEEEbiography}[{\includegraphics[width=1in,height=1.25in,clip,keepaspectratio]{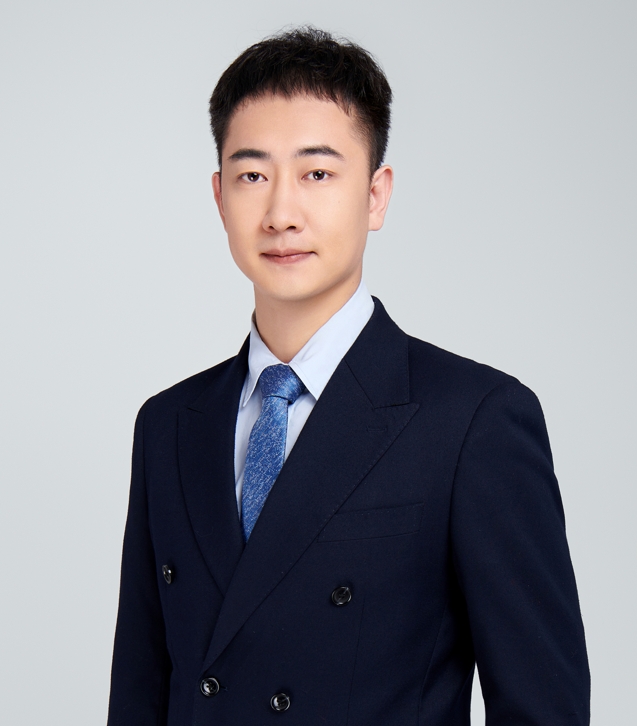}}]
{Sen Chen} (Member, IEEE) is an Associate Professor at the College of Intelligence and Computing, Tianjin University, China. 
Before that, he was a Research Assistant Professor in the School of Computer Science and Engineering, Nanyang Technological University, Singapore.
His research focuses on software security.
He got 6 ACM SIGSOFT Distinguished Paper Awards. More information is available on {\url{https://sen-chen.github.io/}.}
\end{IEEEbiography}

\begin{IEEEbiography}[{\includegraphics[width=1in,height=1.25in,clip,keepaspectratio]{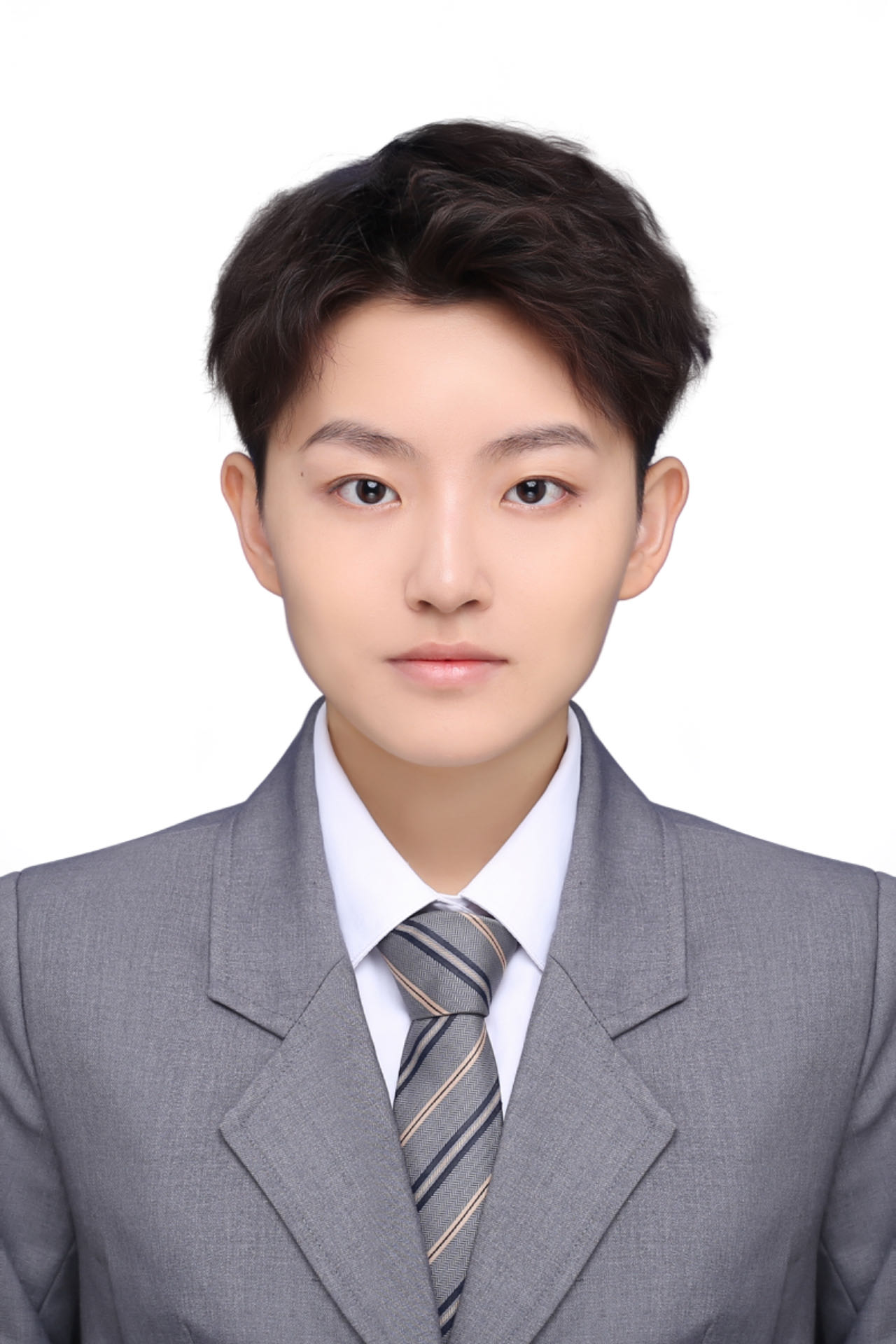}}]
{Miaoying Cai} received her BEng degree in Information Security from Nanjing University of Aeronautics and Astronautics in 2023. She is currently pursuing a Ph.D. degree with the Nankai University. Her research interests lie in the area of mobile security and web security.
\end{IEEEbiography}

\begin{IEEEbiography}
[{\includegraphics[width=1in,height=1.25in,clip,keepaspectratio]{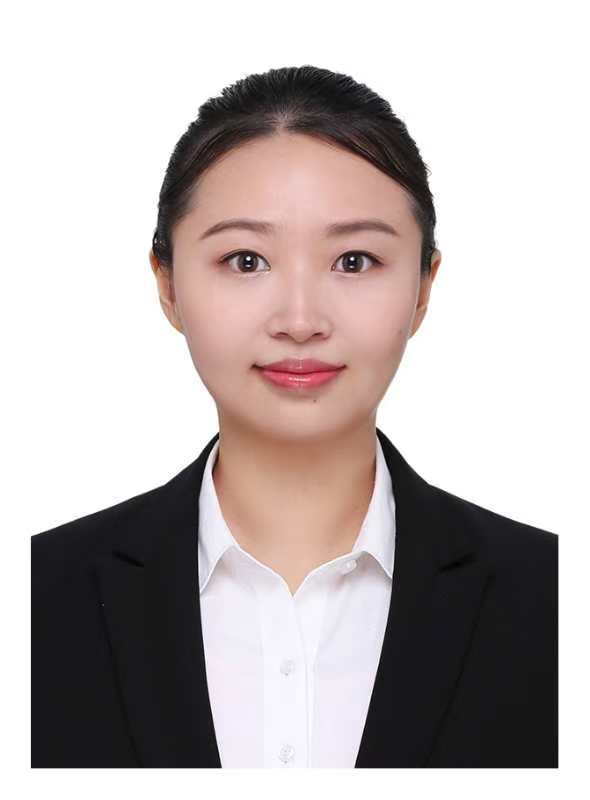}}]
{Sihan Xu} received the B.Sc. and Ph.D. degrees in computer science from Nankai University in 2013 and 2018, respectively. For her research, she spent a year with the National University of Singapore. She is currently an associate professor at the College of Cyber Science, Nankai University. Her research interests include intelligent software engineering and AI security.
\end{IEEEbiography}

\begin{IEEEbiography}
[{\includegraphics[width=1in,height=1.25in,clip,keepaspectratio]{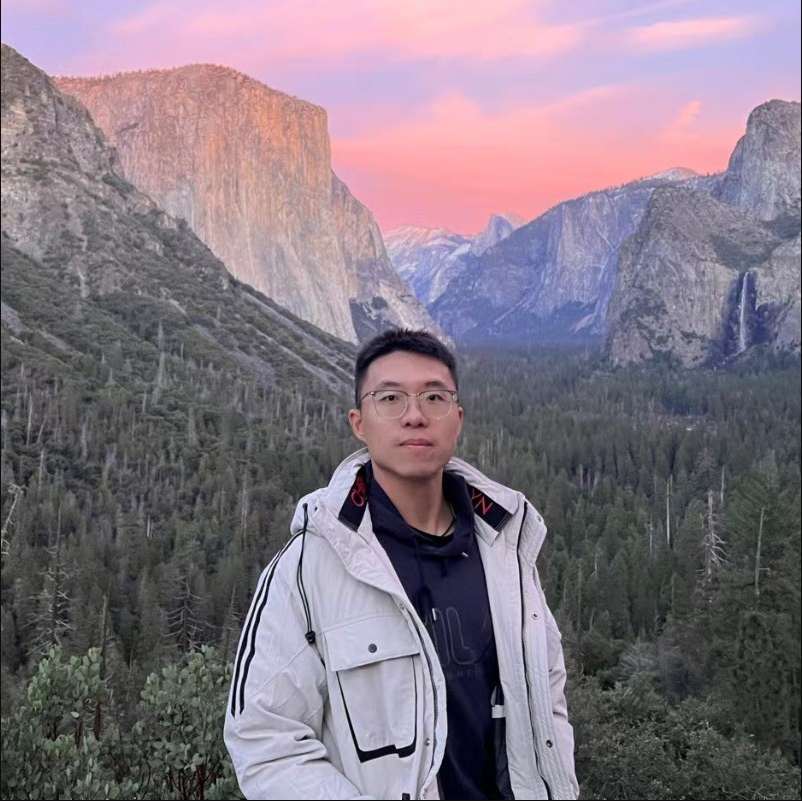}}]
{Lida Zhao} is a Ph.D. candidate at Nanyang Technological University (NTU). His research focuses on software security and software engineering, with a particular emphasis on open-source supply chain security and software composition analysis.
\end{IEEEbiography}

\end{document}